\documentstyle[preprint,aps]{revtex}
\tighten
\draft
\widetext
\input epsf
\preprint{CLNS 98/1567}
\bigskip
\bigskip
\begin{document}
\title{Bose-Fermi Degeneracy and Duality in Non-Supersymmetric Strings}
\medskip
\author{Gary Shiu\footnote{E-mail: shiu@mail.lns.cornell.edu\\
\hbox{$\;\;$} Address after September 1, 1998: Institute for Theoretical 
Physics,
State University of New York \\
\hbox{$\;\;$} at Stony Brook, Stony Brook, NY 11794.} 
and S.-H. Henry
Tye\footnote{E-mail: tye@mail.lns.cornell.edu}}
\bigskip
\address{Newman Laboratory of Nuclear Studies, Cornell University,
Ithaca, NY 14853}
\date{August 16, 1998}
\bigskip
\medskip
\maketitle

\begin{abstract}

{}Following Kachru, Kumar and Silverstein, we construct a set of 
non-supersymmetric Type II string models which 
have equal numbers of bosons and fermions at each mass level.
The models are asymmetric ${\bf Z}_2 \otimes {\bf Z}_2^{\prime}$ orbifolds.
We demonstrate that this bose-fermi degeneracy feature implies 
that both the one-loop
and the two-loop contributions to the cosmological constant vanish. We
conjecture that the cosmological constant
actually vanishes to all loops. We construct a strong-weak dual pair
of models, both of which have bose-fermi degeneracy.
This implies that at least some of the non-perturbative corrections
to the cosmological constant are absent.

\end{abstract}
\pacs{11.25.-w}

\section{Introduction}

{}Bose-fermi degeneracy is well known to be a consequence of supersymmetry.
However, it is not a feature one expects in a non-supersymmetric theory. 
So it is very interesting that, recently, Kachru, Kumar and Silverstein 
\cite{KKS} constructed a $4$-dimensional non-supersymmetric 
Type II string model (the KKS model)
which has this bose-fermi degeneracy feature, {\em i.e.}, the number 
of bosonic and fermionic degrees of freedom are equal at every mass level 
of the spectrum. In this paper, we construct a set of string models that also 
share this same feature.
 
{}The models are non-supersymmetric $4$-dimensional Type II string
models, constructed as 
asymmetric ${\bf Z}_2 \otimes {\bf Z}_2^{\prime}$ orbifolds.
Such models are easy to construct, using the free fermionic string model 
construction \cite{KLT}. 
Starting from the $4$-dimensional ${\cal N}=8$ supersymmetric ({\em i.e.}, 
${\cal N}=(4,4)$) 
Type II model, we introduce a ${\bf Z}_2$ twist which breaks the 
supersymmetry to ${\cal N}=2$ ({\em i.e.}, ${\cal N}=(2,0)$).
On the other hand, we can introduce a different ${\bf Z}_2^{\prime}$ twist
on the original $N=8$ model. 
Judicious choices of these two twists result in 
non-supersymmetric ${\bf Z}_2 \otimes {\bf Z}_2^{\prime}$ orbifold 
models with zero one-loop partition functions.
It turns out that there are two inequivalent types of ${\bf Z}_2^{\prime}$
twist that satisfy the requirements.
The first type of ${\bf Z}_2^{\prime}$ twist breaks supersymmetry to 
${\cal N}=2$ 
({\em i.e.}, ${\cal N}=(0,2)$).
The ${\bf Z}_2^{\prime}$ twist in Ref.~\cite{KKS,harvey} belongs to this
type. 
The other type of ${\bf Z}_2^{\prime}$ twist
breaks supersymmetry to ${\cal N}=4$ 
({\em i.e.}, ${\cal N}=(0,4)$). 
In contrast to the model in 
Ref.~\cite{KKS} where the one-loop partition function
vanishes because of the non-Abelian nature of the orbifold,
the one-loop vanishing in the ${\bf Z}_2 \otimes {\bf Z}_2^{\prime}$
orbifolds presented here is due to the fermionic zero mode.
Another novel feature of this set of models is that there are
massless twisted sector states. Despite the fact that the models
are non-supersymmetric (as the gravitinos are all projected out),
the twisted sector states seem to fall into appropriate 
supermultiplets.

{}The non-trivial question about these models is whether the cosmological
constant remains zero at higher loops. Using an approach similar to that
given in Ref\cite{KKS}, we demonstrate that the cosmological constant 
vanishes at two-loops 
(with the two-loop integrand vanishing point-wise in the moduli space). 
The structure of the models naturally leads to the conjecture that
this property ({\em i.e.}, vanishing multi-loop integrand) persists 
to all loops. We give a plausible argument for this conjecture. 

{}Let us assume that the cosmological constant does vanish to all orders 
in the perturbation expansion. Each model actually represents a class of 
such models, since they have scalar fields that are moduli, {\em i.e.},
the cosmological constant remains zero for different choices of the scalar 
field expectation values. The existence of a set of such models with this 
feature strongly suggests an underlying stringy symmetry that remains to 
be identified. 

{}Some duality aspects of these types of models have been 
explored recently \cite{harvey,KS}. In particular, Kachru and Silverstein
show that the KKS model is self-dual. 
Here we generalize their analysis to more intricate situations.
Some care is necessary to distinguish seemingly similar models 
(with the same orbifold twists and the same counting of states in the
spectra) which
have diferent GSO projections (due to discrete torsions), 
and hence give rise to different spectra. {\em A priori}, 
they have different duality properties, reminiscent of the 
situation of Type IIA/IIB.
This allows us to reproduce the self-duality property of 
the KKS model, and in addition, construct a new
strong-weak dual ({\em i.e.}, $U$-dual) pair,
both of which have bose-fermi degeneracy. 
We argue that
at least some of the non-perturbative corrections
to the cosmological constants are absent.

{}The existence of these models further suggests that it 
may not be too difficult to arrange realistic non-supersymmetric models 
with zero cosmological constant. In 
the realistic universe, the cosmological constant has to vanish only
at an isolated point in the field space, which is a much more relaxed 
constraint. Even if the cosmological constant vanishes in perturbative 
expansion, it may still receive a non-perturbative contribution. At weak 
coupling, we expect at most an exponentially suppressed cosmological 
constant, which may be consistent with nature. 

{}This paper is organized as follows. In Section \ref{prelim}, 
we review the rules 
of the free fermionic string model construction. We follow the notation 
of Ref\cite{KLST}. Readers who are 
familiar with orbifolds but not familiar with 
the fermionic string construction may skip this section, since we will
give 
a translation between the two languages later on.
In Section \ref{Models}, 
we give the explicit constructions of the 
models. It is illuminating to show how the massless spectra emerge under 
the various asymmetric ${\bf Z}_2$ twists. They are shown in the tables.
In Section \ref{Duality}, we show that one of the models constructed 
is self-dual. That model is similar to the KKS model. We also show that 
two of the models form a dual pair. 
In Section \ref{multiloop}, 
we briefly discuss the issues of multi-loop contributions to 
the cosmological constant, and demonstrate the vanishing of the two-loop
vacuum amplitude. We also give a heuristic argument for higher loops.
The last section contains the summary and remarks.

\section{Preliminaries}\label{prelim}

{}In this section we review 
the rules of free fermionic string model construction.
Readers who are famililar with the formulation
may skip this section
and go directly to Section \ref{Models} where the models are presented.
Readers who are familiar with orbifolds but not familiar with
the fermionic string construction may also skip this section, since
a translation between the two languages is given in Section \ref{Models}.
In what follows, we will concentrate on Type II string vacua with
four non-compact space-time dimensions in the light-cone RNS formulation.
In the free-fermionic construction,
all of the world-sheet degrees of freedom corresponding to the compactified 
six space-like dimensions are fermionized. Moreover, these degrees of 
freedom are described via {\em free} world-sheet fermions, which are written
in the complex basis, with various spin structures, that is, 
boundary conditions.

{}We have the following world-sheet degrees of freedom: one complex 
world-sheet boson $\phi^0$ corresponding to two real transverse space-time 
coordinates in the light-cone gauge;    
one complex fermion 
$\psi^0$ which is the world-sheet superpartner of $\phi^0$; 
three pairs of complex fermions $\chi^\ell$ and $\lambda^\ell$, $\ell=1,2,3$, 
corresponding to fermionization of three complex (six real) world-sheet 
bosons $\phi^\ell$ describing the compactified dimensions; three complex 
world-sheet fermions $\psi^\ell$ which are superpartners of the world-sheet 
bosons $\phi^\ell$. Since we are discussing Type II strings, each of the 
above world-sheet degrees of freedom has a left- and right-moving component 
which we will denote by subscripts ``$L$'' and ``$R$'', respectively.

{}In the following discussion we will mostly deal with the world-sheet 
fermions $\psi^0,\psi^\ell,\chi^\ell,\lambda^\ell$. We will collectively 
refer to them as $\Psi^r$, $r=0,\dots,9$, where $\Psi^0\equiv\psi^0$, 
$\Psi^{\ell}\equiv\psi^\ell$, $\Psi^{1+\ell}\equiv
\chi^\ell$, and $\Psi^{2+\ell}\equiv\lambda^\ell$. Similar notation will 
be used for the corresponding left- and right-moving components as well.

{}It is convenient to organize the string states into sectors labeled by 
the monodromies of the world-sheet degrees of freedom. Thus, consider the 
sector where
\begin{eqnarray}
 &&\Psi^r_L (ze^{2\pi i}) = \exp(-2\pi i v_{Li})\Psi^r_L (z)~,\\
 &&\Psi^r_R ({\overline z}e^{-2\pi i}) 
= \exp(-2\pi i v_{Ri})\Psi^r_R ({\overline z})~.
\end{eqnarray}  
Note that 
$\phi^0(ze^{2\pi i},{\overline z}e^{-2\pi i})=\phi^0(z,{\overline z})$ 
since $\phi^0$ corresponds to space-time coordinates. These monodromies 
can be combined into a single vector
\begin{equation}
 V_i=\left[v^0_{Li}(v^1_{Li}v^2_{Li}v^3_{Li})(v^4_{Li}v^5_{Li}v^6_{Li})(v^7_{Li}v^8_{Li}v^9_{Li})
 || v^0_{Ri}(v^1_{Ri}v^2_{Ri}v^3_{Ri})(v^4_{Ri}v^5_{Ri}v^6_{Ri})
 (v^7_{Ri}v^8_{Ri}v^9_{Ri})\right]~.
\end{equation}
The double vertical line separates the monodromies corresponding to the 
left- and right-moving components. Without loss of generality we can 
restrict the values of $v^r_{Li},v^r_{Ri}$ as 
follows: $-{1\over 2} \leq v^r_{Li},v^r_{Ri} <{1\over 2}$. Note that 
if a given monodromy is $-{1\over 2}$ then the corresponding world-sheet 
degree of freedom is a complex Ramond fermion; if this monodromy is 0, 
then it is a Neveu-Schwarz complex fermion.

{}The monodromies $V_i$ can be viewed as fields $\Psi^r$ being periodic 
$\Psi^r(ze^{2\pi i},{\overline z}e^{-2\pi i})=\Psi^r(z,{\overline z})$ 
up to the identification $\Psi^r\sim g(V_i) \Psi^r
g^{-1} (V_i)$, where $g(V_i)$ is an element of a finite 
discrete {\em orbifold} group $\Gamma$. In this paper we 
will only consider Abelian orbifold groups $\Gamma$. In fact, 
we will focus on orbifold groups which are direct products of 
${\bf Z}_2$ subgroups. In these cases all the $V_i$ vectors only 
have elements taking values 0 or $-{1\over 2}$.

{}The requirement of world-sheet supersymmetry is 
necessary to ensure space-time Lorentz invariance in the covariant gauge, 
which implies that the world-sheet supercurrents must have well defined
monodromies. That is, for $\ell=1,2,3$:
\begin{eqnarray}\label{triplet}
 &&v^{\ell}_{Li}+v^{\ell+1}_{Li}+v^{\ell+2}_{Li}\equiv s_i~({\mbox{mod}}~1)~,\\
 &&v^{\ell}_{Ri}+v^{\ell+1}_{Ri}+v^{\ell+2}_{Ri}\equiv {\overline s}_i~({\mbox{\
mod}}~1)~.
\end{eqnarray}
where $s_i\equiv v^0_{Li}$ and ${\overline s}_i\equiv v^0_{Ri}$ 
determine whether the corresponding space-time states are bosons 
or fermions: the NS-NS sectors with $s_i={\overline s}_i=0$ as well as 
the R-R sectors with 
$s_i={\overline s}_i=-{1\over 2}$ give rise to space-time bosons; the 
NS-R sectors with 
$s_i=0, {\overline s}_i=-{1\over 2}$ as well as the R-NS sectors with 
$s_i=-{1\over 2}, {\overline s}_i=0$ give rise to space-time fermions. 

{}The notation we have introduced proves convenient in describing the 
sectors of a given string model with an orbifold group $\Gamma$. As we 
have already mentioned, we will confine our attention to the orbifold 
groups $\Gamma\approx({\bf Z}_2)^{\otimes n}$ so that all the elements 
$V_i^s$ ($s=0,\dots, 19$) are either 0 or $-{1\over 2}$. 
Here $V^s_i=v^r_{Li}$ for $s=r=0,\dots,9$, and $V^s_i=v^r_{Ri}$ 
for $s=r+10=10,\dots,19$. To describe all of the $2^n$ elements 
of group $\Gamma$, it is convenient to introduce a set of generating 
vectors $\{V_i \}$ such that ${\overline{\alpha V}}={\bf 0}$ if and 
only if $\alpha_i\equiv 0$. Here ${\bf 0}$ is the null vector:
\begin{equation}
 {\bf 0}  =\left[0(000)(000)(000) || 0(000)(000) (000)\right]\equiv 
\left[0(000)^3 || 0(000)^3\right]~.
\end{equation}
Also, $\alpha V\equiv \sum_i \alpha_i V_i$ with the summation defined 
as $(V_i+V_j)^s=
V^s_i + V^s_j$, and $\alpha_i=0,1$. The overbar notation is defined as 
${\overline {\alpha V}}\equiv \alpha V- \Delta (\alpha)$, and the elements 
of ${\overline {\alpha V}}$ satisfy $-{1\over 2}\leq  {\overline {\alpha V}}^s <{1\over 2}$, where $\Delta^s (\alpha)\in {\bf Z}$. The elements $g({\overline {\alpha V}})$ of the group $\Gamma$ are in one-to-one correspondence with the vectors ${\overline {\alpha V}}$. It is the Abelian nature of $\Gamma$ that allows for this correspondence by simply taking all possible linear combinations of the generating vectors $V_i$. 

{}Now we can identify the sectors of a given model. They are labeled by 
the vectors 
${\overline {\alpha V}}$, and in a given sector ${\overline {\alpha V}}$ the monodromies of the string degrees of freedom are given by $\Psi^r (ze^{2\pi i},{\overline z}e^{-2\pi i})=
g({\overline {\alpha V}}) \Psi^r (z,{\overline z}) g^{-1}({\overline {\alpha V}})$. The sectors with 
${\overline {\alpha V}}^0={\overline {\alpha V}}^{10}$ give rise to space-time bosons, whereas the sectors with ${\overline {\alpha V}}^0\not={\overline {\alpha V}}^{10}$ give rise to space-time
fermions.

{}In the next section we will show that the following vector must always be present among the generating vectors:
\begin{equation}
 V_0  = \textstyle{
\left[-{1\over 2}\left(-{1\over 2}-{1\over 2}-{1\over 2}\right)^3 || 
 -{1\over 2}\left(-{1\over 2}-{1\over 2}-{1\over 2}\right)^3\right]}~.
\end{equation}
(The presence of this vector is required by one-loop modular invariance.) In fact, the 
$\Gamma\approx {\bf Z}_2$ orbifold model (which contains two sectors ${\bf 0}$ and $V_0$)
is a non-supersymmetric theory without any fermions in its spectrum. To construct supersymmetric theories we must add additional vectors to the generating set $\{V_i \}$. Consider adding two more vectors of the following form:
\begin{eqnarray}
 &&V_1  =\textstyle{\left[-{1\over 2}\left(-{1\over 2}00\right)^3|| 
 0(000)^3\right]}~,\\
 &&V_2  =\textstyle{
\left[  0(000)^3|| -{1\over 2}\left(-{1\over 2}00\right)^3
\right]}~.
\end{eqnarray}
The orbifold group now is $\Gamma\approx ({\bf Z}_2)^{\otimes 3}$. Such a 
model (subject to the consistency conditions discussed in the subsequent 
sections) corresponds to a four dimensional Type IIA or Type IIB string 
theory (depending on certain phases entering the one-loop partition 
function - see below) with ${\cal N}=8$ space-time supersymmetry.

{}As mentioned before, we will focus on orbifold groups that are 
direct products of ${\bf Z}_2$ subgroups. Hence the boundary conditions
of the complex fermions $\Psi^{\ell}$
can only take values $0$ or $1/2$. 
This implies that the worldsheet fermions do not need to be written in
the complex basis (in contrast to other ${\bf Z}_N$ orbifolds where
the monodromies are complex and so a complex basis of the fermions
is necessary). 
In cases where the worldsheet degrees of freedom contain
real fermions, the generating set $\{V_i \}$ 
must satisfy the {\em cubic} constraint \cite{KLST}:
\begin{equation}\label{cubic}
4 \sum_{\ell: real} V^{\ell}_i V^{\ell}_j V^{\ell}_k = 0 \pmod{1}
~~\mbox{for all}~i,j,k~.
\end{equation}
This ensures that we can find a complex basis for any three generating
vectors, even though there is no complex basis which is common
for {\em all} the generating vectors.

\subsection{One-Loop Modular Invariance}

{}The $g$-loop scattering amplitudes must satisfy modular invariance.
Let us start with the one-loop vacuum amplitude: a torus. 
Conformally inequivalent tori are labeled by a single complex modular 
parameter $\tau=\tau_1+i\tau_2$.
A torus has two non-contractable cycles $a$ and $b$ depicted as follows:
\begin{center}
\hspace{1cm}
\epsfxsize=7 cm
\epsfbox{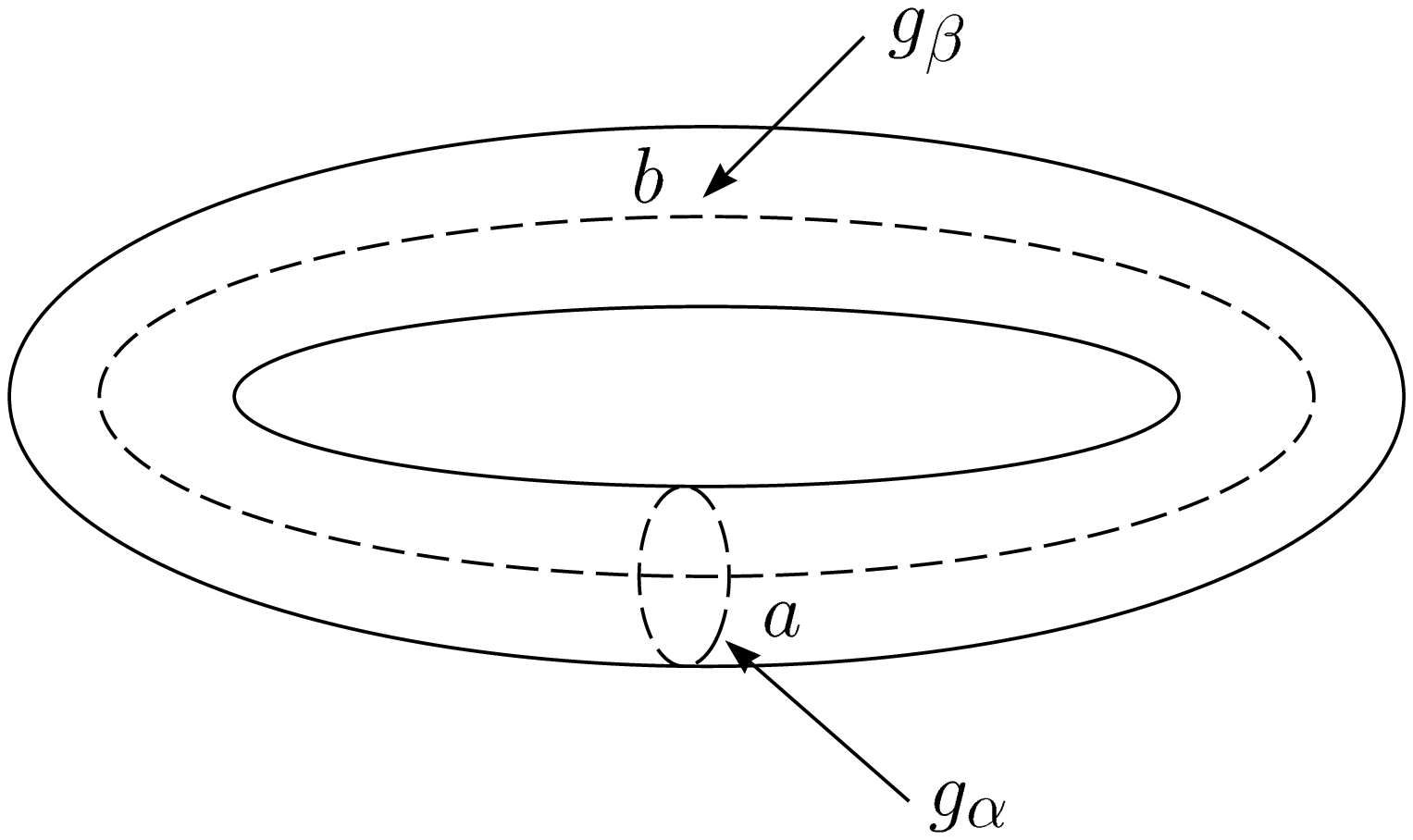}
\end{center}

{}The identification for the $a$ cycle corresponds to considering states 
from the $g_\alpha$ sector propagating in the loop. 
The identification for the $b$ cycle corresponds to inserting the 
$g_\beta$ twist into the path integral for the world-sheet fermions. 
Thus, the world-sheet fermionic contribution to the 
corresponding vacuum amplitude $Z^\alpha_\beta$ is given by: 
\begin{equation}
 Z^\alpha_\beta={\mbox{Tr}}\left(\exp\left[2\pi i \tau H_\alpha -2\pi i{\overline\tau}
 {\overline H}_\alpha\right] g_\beta\right)~,
\end{equation}  
where $H_\alpha$ and ${\overline H}_\alpha$ are the Hamiltonians 
corresponding to the left- and right-moving fermions, and the trace is 
taken over all the states in the $g_\alpha$ twisted sector. 
The character $Z^\alpha_\beta$ is simply given by a product of 
characters $Z^{{\overline {\alpha V}}^s}_{{\overline {\beta V}}^s}$ for each world-sheet fermion:
\begin{equation}\label{ind}
 Z^\alpha_\beta=\prod_{s=0}^{19} Z^{{\overline {\alpha V}}^s}_{{\overline {\beta V}}^s}~,
\end{equation} 
where we have taken into account that the action of $g_\beta$ on any given state in the $g_\alpha$ twisted sector amounts to simply multiplying it by a phase. This follows from the fact that here we are considering an Abelian orbifold. The individual characters $Z^{{\overline {\alpha V}}^s}_{{\overline {\beta V}}^s}$ are given by (up to overall phases):
\begin{equation}\label{char}
 Z^v_u=\left[\eta(\tau)\right]^{-1} \sum_{n\in {\bf Z}} \exp\left(i\pi (n-v)^2 \tau +2\pi i n u\right)~,
\end{equation}
for a left-moving fermion with boundary conditions $v$ and $u$ around the $a$ and $b$ cycles, and by the complex conjugate of this expression for a right-handed fermion with the same boundary conditions. Here $\eta(\tau)$ is the Dedekind $\eta$-function.

{}The world-sheet fermionic contribution to the one-loop partition function is obtained by summing over all possible boundary conditions $\alpha$ and $\beta$ around the $a$ and $b$ cycles:
\begin{equation}\label{part}
 Z_1 =\sum_{\alpha,\beta} C^\alpha_\beta Z^\alpha_\beta~,
\end{equation} 
where $C^\alpha_\beta$ are certain phases which we have omitted in 
(\ref{char}), so that $Z^\alpha_\beta$ in (\ref{part}) are computed 
using (\ref{ind}) with the characters 
$Z^{{\overline {\alpha V}}^s}_{{\overline {\beta V}}^s}$ defined as 
in (\ref{char}) {\em without} the phases. 
Consistency requires $Z_1$ to be modular invariant, {\em i.e.}, invariant 
under the modular transformations generated by $\tau\rightarrow\tau+1$ 
and $\tau\rightarrow-{1\over\tau}$. 

{}Thus, let us consider the behavior of $Z^{\overline {\alpha V}}_{\overline {\beta V}}
\equiv Z^\alpha_\beta$ under the modular transformations:
\begin{eqnarray}
 &&\tau\rightarrow -{1\over \tau}:~~~Z^{\overline {\alpha V}}_{\overline {\beta V}}
 \rightarrow \exp\left(2\pi i {\overline {\alpha V}}\cdot {\overline {\beta V}}\right)
 Z^{\overline {\beta V}}_{\overline {-\alpha V}}~,\\
 &&\tau\rightarrow \tau+1:~~~Z^{\overline {\alpha V}}_{\overline {\beta V}}
 \rightarrow \exp\left(\pi i  {\overline {\alpha V}}\cdot {\overline {\alpha V}}
 \right)
 Z^{\overline {\alpha V}}_{\overline {\beta V-\alpha V+V_0}}~,
\end{eqnarray}
where the dot product of two vectors is defined as follows:
\begin{equation}
 {\overline {\alpha V}}\cdot {\overline {\beta V}}\equiv \sum_{s=0}^9 
 {\overline {\alpha V}}^s {\overline {\beta V}}^s -
 \sum_{s=10}^{19} 
 {\overline {\alpha V}}^s {\overline {\beta V}}^s~,
\end{equation}
that is, this dot product is defined with the Lorentzian signature $((+)^{10},(-)^{10})$ with the plus and minus signs corresponding to the left- and right-moving components, respectively. To ensure that $Z_1$ is modular invariant, we must require that 
\begin{eqnarray}\label{modS}
 &&\tau\rightarrow -{1\over \tau}:~~~C^{\overline {\alpha V}}_{\overline {\beta V}}
 \exp\left(2\pi i {\overline {\alpha V}}\cdot {\overline {\beta V}}\right) =
 C^{\overline {\beta V}}_{\overline {-\alpha V}}~,\\
\label{modT}
 &&\tau\rightarrow \tau+1:~~~C^{\overline {\alpha V}}_{\overline {\beta V}}
 \exp\left(\pi i {\overline {\alpha V}}\cdot {\overline {\alpha V}}\right) =
 C^{\overline {\alpha V}}_{\overline {\beta V-\alpha V+V_0}}~,
\end{eqnarray}

\subsection{Multi-loop Modular Invariance and Factorization}

{}The requirements of multi-loop modular invariance and factorization 
impose additional constraints on these phases. 
For a given generating set of vectors $\{V_i\}$, which must satisfy 
certain constraints, this allows to solve for these phases in terms of 
certain discrete constants.

{}The corresponding $g$-loop $Z_g$ reads:

\begin{equation}\label{partg}
 Z_g =\sum_{\alpha_i,\beta_i}  \zeta_g 
C^{\alpha_1,\dots,\alpha_g}_{\beta_1,\dots,\beta_g}
 Z^{\alpha_1,\dots,\alpha_g}_{\beta_1,\dots,\beta_g}~,
\end{equation}         
where $\zeta_g(\tau)$ is a factor that includes time-like and logitudinal 
modes, (super-)ghost modes and picture-changing operator insertions.
$Z^{\alpha_1,\dots,\alpha_g}_{\beta_1,\dots,\beta_g}$ is given by 
a product of the contributions corresponding to individual left- and 
right-moving complex fermions. The characters for a left-moving fermion read:
\begin{equation}\label{charg}
 Z^{v_1,\dots,v_g}_{u_1,\dots,u_g}=D_g(\tau) \sum_{\{n_i\}\in 
{\bf Z}^{\otimes g}} 
 \exp\left(i\pi (n_i-v_i)(n_j-v_j)\tau_{ij} +2\pi i n_i u_i\right)~,
\end{equation}
where $D_g(\tau)$ is a certain function of $\tau_{ij}$ whose precise 
form is not going to be important in the following. 

{}From (\ref{charg}) we see that the characters $Z^{\alpha_1,\dots,\alpha_g}_{\beta_1,\dots,\beta_g}$ indeed factorize:
\begin{equation}\label{fact}
 Z^{\alpha_1,\dots,\alpha_g}_{\beta_1,\dots,\beta_g} (\tau,{\overline{\tau}}) \rightarrow\prod_i
 Z^{\alpha_i}_{\beta_i} (\tau_{ii},{\overline{\tau}})~.
\end{equation}
Thus, for $Z_g$ to factorize we must require that the phases satisfy the following condition:
\begin{equation}\label{phases}
 C^{\alpha_1,\dots,\alpha_g}_{\beta_1,\dots,\beta_g}\equiv \prod_i
 C^{\alpha_i}_{\beta_i}~.
\end{equation}

{}Next, let us consider requirements that higher genus modular invariance 
imposes on the phases $C^\alpha_\beta$. It fact, it is is sufficient to 
consider these constraints at $g=2$ as they are powerful enough to solve 
for the phases $C^\alpha_\beta$. Once these phases are obtained 
using the genus-two constraints, it is straightforward to check that 
higher genus constraints are also satisfied.
At genus two we have  
\begin{equation}\label{part2}
 Z_2 =\sum_{\alpha_i,\beta_i} \zeta_2 C^{\alpha_1}_{\beta_1} 
C^{\alpha_2}_{\beta_2}  Z^{\alpha_1,\alpha_2}_{\beta_1,\beta2}~.
\end{equation}
Now consider the $Sp(4,{\bf Z})$ modular transformation $\tau_{12}\rightarrow\tau_{12}+1$.
The corresponding constraint reads:
\begin{equation}\label{mod2}
 C^{\overline{\alpha V}}_{\overline{\beta V}} 
 C^{\overline{\gamma V}}_{\overline{\delta V}}\exp\left(2\pi i
 \left[{\overline{\alpha V}}\cdot{\overline{\gamma V}}+\alpha(s+{\overline s}) +
 \gamma(s+{\overline s})\right]\right)
=C^{\overline{\alpha V}}_{\overline{(\beta - \gamma) V}} 
 C^{\overline{\gamma V}}_{\overline{(\delta - \alpha) V}} ~.
\end{equation}
Here the phase $2\pi i{\overline{\alpha V}}\cdot{\overline{\gamma V}}$ comes from the modular
transformation of the character $Z_g$ corresponding to $\Psi^r$. However, the phase $2\pi i
\left[\alpha(s+{\overline s}) +\gamma(s+{\overline s})\right]$ comes from 
the modular transformation of $\zeta_2$.

{}The two-loop modular invariance constraints (\ref{mod2}) together with the one-loop modular invariance constraints (\ref{modS}) plus (\ref{modT}) are restrictive enough to solve for the phases $C^{\overline{\alpha V}}_{\overline{\beta V}}$. The most general solution reads:
\begin{equation}\label{C}
 C^{\overline{\alpha V}}_{\overline{\beta V}}=\exp\left[2\pi i\beta\phi(\alpha)+\alpha (s+
 {\overline s})\right]~,
\end{equation}
where
\begin{equation}\label{phi}
 \phi_i(\alpha)\equiv\sum_j k_{ij}\alpha_j + s_i +{\overline s}_j -V_i\cdot {\overline{\alpha V}}~,
\end{equation}
and the {\em structure constants} $k_{ij}=0,1/2$ must satisfy the following constraints:
\begin{eqnarray}
 &&k_{ij}+k_{ji}-V_i\cdot V_j =0~({\mbox{mod}}~1)~,\\
 &&k_{ii}+k_{i0}+s_i+{\overline s}_i-{1\over 2} V_i\cdot V_i=0 ~({\mbox{mod}}~1)~.
\end{eqnarray}
These constraints together with (\ref{C}) and (\ref{phi}) are necessary and sufficient to guarantee multi-loop modular invariance. Here the factorized form of (\ref{phases}) is understood.

\section{Non-supersymmetric Free Fermionic String Models and One-loop
Cosmological Constant}\label{Models}

{}In this section, we present a class of non-supersymmetric
string models constructed in the free fermionic framework.
In spite of the fact that the models are non-supersymmetric, 
there is an equal number of bosons and fermions at each mass level.
This implies that not just
the one-loop contribution to the cosmological constant $\Lambda$
is zero, but that the one-loop amplitude vanishes point by point
in the moduli-space of Riemann surfaces . 
In constrast to Ref. \cite{KKS} in which the one-loop
contribution vanishes because of the non-Abelian nature of the orbifold,
the one-loop vanishing in this class of models is due to the
fermionic zero modes.

{}Let us start with Type II string theory compactified on $T^6$. 
As discussed before, the vectors $V_1$ and $V_2$ are always present 
in the generating set $\{ V_i \}$. Therefore, before orbifolding,
the model has ${\cal N}=(4,4)$ supersymmetry.
Let us denote the
corresponding one-loop partition function by $Z_{\bf 0}$.
Consider orbifolding Type II theory by two {\em commuting}
${\bf Z}_2$ elements $f$ and $g$, the one-loop partition function is
given by
\begin{equation}
Z = {1\over 4} \sum_{k,l,m,n} Z(f^k g^l, f^m g^n)
\end{equation}
where $f^k g^l$ and $f^m g^n$ are the twists in the $a$- and $b$-cycle
respectively. 

{}Suppose that the term $Z(f,g)$ vanishes, then by one-loop
modular transformation,
the terms $Z(f,f^m g)$, $Z(g, f g^m)$, $Z(f^k g, f)$ and $Z(f g^l, g)$
also vanish. The expression can then be simplified to:
\begin{equation}\label{sum}
Z = {1\over 2} Z_f + {1\over 2} Z_g
   +{1\over 2} Z_{fg} - {1\over 2} Z_{\bf 0}
\end{equation}
where $Z_f$, $Z_g$ and $Z_{fg}$ are the one-loop partition
functions of the ${\bf Z}_2$ orbifolds generated by $f$, $g$ 
and $fg$ respectively. 
Clearly, $Z_{\bf 0}$ is zero because of supersymmetry.
If the individual ${\bf Z}_2$ orbifolds (generated by $f$, $g$ and $fg$
respectively) are supersymmetric, then the total one-loop
partition function is identically zero. So the strategy is to find a pair 
of twists $f$ and $g$ with the above properties. 

{}The generating vectors that are already present in the
original ${\cal N}=8$ model are given by:
\begin{eqnarray}
 V_0  &=& \textstyle{
\left[-{1\over 2}\left(-{1\over 2}-{1\over 2}-{1\over 2}\right)^3 || 
 -{1\over 2}\left(-{1\over 2}-{1\over 2}-{1\over 2}\right)^3\right]}~, \\
 V_1  &=&\textstyle{\left[-{1\over 2}\left(-{1\over 2}00\right)^3|| 
 0(000)^3\right]}~,\\
 V_2  &=&\textstyle{
\left[  0(000)^3|| -{1\over 2}\left(-{1\over 2}00\right)^3
\right]} ~.
\end{eqnarray}
The 
${\bf Z}_2$ elements $f$ and $g$ can be represented by the
additional generating vectors $V_3$ and $V_4$. The constraints from
multi-loop modular invariance discussed in the previous section
imply:

\begin{eqnarray}
(i) 
\hspace{5cm} V_i \cdot V_1 &=& 0,1/2 \pmod{1} ~, 
\hspace{3cm} \nonumber\\
     \hspace{5cm} V_i \cdot V_2 &=& 0,1/2 \pmod{1} ~, ~~\mbox{for i=3,4}~.
\hspace{3cm} \nonumber\\
&& \nonumber \\
(ii) \hspace{5cm} V_3^2, V_4^2 &\in& {\bf Z}~.   \nonumber \\
&& \nonumber \\
(iii) \hspace{5cm} V_3 \cdot V_4 &=& 0, 1/2 \pmod{1}~. \nonumber
\end{eqnarray}



{}Without loss of generality, we can take $s_i=\overline{s}_i=0$.
This is because $V_1$ and $V_2$ are among the generating vectors,
and adding these vectors changes the spin statistics.
The supercurrent constraint (Eq.(\ref{triplet})) and ($i$) restrict
the form of $V_i=(v_{Li} \vert \vert v_{Ri})$. Let us consider
$v_{Li}$ for the moment (as the analysis for $v_{Ri}$ is completely
parallel), it takes one of the following forms:
\begin{eqnarray}
V_i \cdot V_1 = 0: \quad \quad \quad
v_{Li} &=& \textstyle{\left[ 0 ~(0 ~a ~a ) (0 ~b ~b) (0 ~c ~c) \right]}~. \\
V_i \cdot V_1 = {1\over 2}: \quad \quad \quad 
v_{Li} &=& \textstyle{\left[ 0 \left( -{1\over 2} -{1\over 2} ~0 \right)^2
                               (0 ~d ~d) \right]} ~.
\end{eqnarray}
where $a,b,c,d=0$ or $1/2$. 

{}One can also include in the orbifold twist ($f$ and $g$) 
an action of $(-1)^{F_L}$ which acts with a
$(-1)$ on all spacetime spinors coming from the left-moving degrees
of freedom. This is equivalent to turning on a discrete torsion \cite{Torsion}
since in the left-moving Ramond sector the orbifold
projection has the opposite sign from what it would have without the
$(-1)^{F_L}$ action. The discrete torsion is determined by the 
choice of the {\em structure constants} $k_{1i}$ which can take values
0 or $1/2$.
Choosing $k_{1i}=1/2$ is equivalent to including the $(-1)^{F_L}$ 
action in the orbifold twist. Similarly, one can include the $(-1)^{F_R}$
action in the orbifold twist by choosing $k_{2i}=1/2$.

{}In the case where $V_i \cdot V_1=0$,
the number of supersymmetries remains after orbifolding
depends on the discrete torsion.
The $4$ gravitinos in the R-NS sector remain in the
spectrum if $k_{1i}=0$ (hence ${\cal N}_L=4$) 
but are projected out if $k_{1i}=1/2$ (hence ${\cal N}_L=0$). 
For $V_i \cdot V_1 =1/2$, the number of supersymmetries broken by the
orbifold twist is independent of the discrete torsion.
The choice of $k_{1i}$
simply determines which two gravitinos are projected out.
The number of supersymmetries remains is ${\cal N}_L=2$.

{}The individual ${\bf Z}_2$ orbifolds (generated by $f$, $g$ and $fg$)
are supersymmetric whereas the resulting ${\bf Z}_2 \otimes {\bf Z}_2^{\prime}$
orbifold is non-supersymmetric. By counting the number
of gravitinos, it is easy to see from
Eq.(\ref{sum}) that there are two independent classes of models.
In the first class of models, the orbifold twists $f$, $g$ and $fg$ break 
${\cal N}=(4,4)$ 
supersymmetry to ${\cal N}=(2,0),(0,2)$ and $(2,2)$ respectively.
The model presented in Ref. \cite{KKS} (and also its variant in 
Ref. \cite{harvey}) belongs to this class. 
In the other class of models, the orbifold twists $f$, $g$ and $fg$ break 
${\cal N}=(4,4)$ 
supersymmetry to ${\cal N}=(2,0),(0,4)$ and $(2,0)$ respectively.

{}Before we consider the non-supersymmetric 
models in detail, let us briefly
discuss the massless spectrum
of the original ${\cal N}=8$ supersymmetric model to
set up our notation.
There are $4$ sectors that give rise to massless states:
${\bf 0}$ sector ({\em i.e.}, NS-NS sector) and $V_1 + V_2$ sector
({\em i.e.}, R-R sector) 
give rise to spacetime bosons, $V_1$ sector ({\em i.e.}, R-NS sector) and
$V_2$ sector ({\em i.e.}, NS-R sector) give rise to spacetime fermions.
The graviton $G_{ij}$, the antisymmetic tensor $B_{ij}$ and
the dilaton $\phi$ come from the NS-NS sector. In addition, 
there are $U(1)^6 \otimes U(1)^6$ gauge bosons and $36$ real scalars
in this sector 
due to the compactification on $T^6$. Therefore the NS-NS sector
has $64$ bosonic degrees of freedom.
The R-R sector also provides $U(1)$ gauge fields. They differ from
that in the NS-NS sector in that they are obtained
by tensoring the left- and right-moving spinor states. This gives rise to
$U(1)^{16}$ gauge bosons and $32$ scalars, a total of $64$ bosonic
degrees of freedom. Because of the ${\cal N}_L=4$ supersymmetry,
there are $4$ gravitinos in the R-NS sector. Together with the
$28$ spinors in this sector, they provide $64$ fermionic degrees of
freedom. The spectrum in the NS-R sector is similar to that in the
R-NS sector (with the left- and right-moving quantum numbers interchanged).
Hence it also gives rise to $64$ fermionic degrees of freedom.
The one-loop partition function $Z_{{\bf 0}}$ vanishes because
of supersymmetry. Therefore, 
the number of bosonic and fermionic degrees of freedom
are equal at each mass level. In particular, we have seen that this is 
the case at the massless level.

{}Let us now turn to the non-supersymmetric models:

\subsection{Class I}


{}The general forms of $V_3$ and $V_4$ are given by:
\begin{eqnarray}
V_3 &=& \textstyle{\left[ 0 \left( -{1\over 2} -{1\over 2} ~0 \right)^2
                               (0 ~d_{3} ~d_{3}) 
      \vert \vert  0 ~(0 ~a_3 ~a_3 ) (0 ~b_3 ~b_3) (0 ~c_3 ~c_3) 
\right]}~, \\
V_4 &=& \textstyle{\left[ 0 ~(0 ~a_4 ~a_4 ) (0 ~b_4 ~b_4) (0 ~c_4 ~c_4) 
\vert \vert
0 \left( -{1\over 2} -{1\over 2} ~0 \right)^2
                               (0 ~d_{4} ~d_{4}) \right]} ~. 
\end{eqnarray}
The structure constants $k_{23}=1/2$, $k_{14}=1/2$, whereas the other
$k_{ij}$ are not fixed by the supersymmetries of the individual
${\bf Z}_2$ orbifolds (these $k_{ij}$ determine
which two of the gravitinos are projected out and which two are kept.
Each choice corresponds to a different GSO projection).
This implies that with the same orbifold twists, there
exists more than one model. This point will be crucial later on
when we discuss the issue of duality.
Here, $a_i,b_i,c_i$ and $d_i$ are chosen so that $V_i^2 \in {\bf Z}$
and $V_3 \cdot V_4 = 0~\mbox{or}~ 1/2 ~(\mbox{mod} ~1)$.
The one-loop partition function of the ${\bf Z}_2 \otimes {\bf Z}_2^{\prime}$
orbifold is zero if we demand the contribution of the one-loop
diagram with $f$ twist on the $a$-cycle and $g$ twist 
on the $b$-cycle (for {\em all} space-time spin structure) to vanish.
In other words,
\begin{equation}
Z (V_3 + \sum_{i=0}^2 \alpha_i V_i,
V_4 + \sum_{i=0}^2 \beta_i V_i) = 0
\end{equation} 
for all $\alpha_i$ 
and $\beta_i$. This can be achieved by demanding that 
at least one of the $20$ complex worldsheet fermions 
($10$ left-moving and $10$ right-moving) has periodic
boundary conditions in both the $a$-cycle and the $b$-cycle.
In other words,
\begin{equation}\label{cross}
V^a_3 + \sum_{i=0}^{2} \alpha_i V^a_i 
= V^a_4 + \sum_{i=0}^2 \beta_i V^a_i = {1 \over 2}~,
~~{\mbox{for some}}~ a=1,2,\dots,20
\end{equation}
for all $\alpha_i$ and $\beta_i$.
Then the one-loop diagram is proportional to $\theta \left[ 1/2, 1/2 \right]$
which vanishes because of the fermionic zero mode.

{}There are many choices of $a_i,b_i,c_i$ and $d_i$ which
satisfy the constraints, but not all choices are independent
(since $V_0$, $V_1$ and $V_2$ are always in the generating set, adding
linear combinations of these vectors to $V_3$ and $V_4$ gives
rise to a seemingly different but equivalent set of generating
vectors).
It turns out there are two models in this class (plus their
variations): \\

$\bullet$ \underline{Model IA} 
\begin{eqnarray}
V_3 &=& \textstyle{\left[ 0 ~( -{1\over 2} -{1\over 2} ~0 )^2
                               (0 0 0) 
      \vert \vert  0 ~(0 -{1\over 2} -{1\over 2})^2 (0 0 0) \right]}~, \\
V_4 &=& \textstyle{\left[ 0 ~(0 -{1\over 2} -{1\over 2})^2 (0 0 0) 
\vert \vert
0 ~( -{1\over 2} -{1\over 2} ~0 )^2
                               (0 0 0) \right]} ~. 
\end{eqnarray}

{}The twists $f$ and $g$ are left-right mirror of each other, {\em i.e.}
$v_{3L}=v_{4R}$ and $v_{3R}=v_{4L}$. Therefore, the spectrum of the
${\bf Z}_2$ orbifold generated by $f$ can be obtained from that of $g$ 
by interchanging 
the left- and right-moving quantum numbers. In what follows, the 
${\bf Z}_2$ orbifold refers to the orbifold generated by $f$.

 
{}The untwisted NS-NS sector (${\bf 0}$ sector)
gives rise to the graviton $G_{ij}$, the 
antisymmetric tensor $B_{ij}$ and the dilaton $\phi$; all of them survive
both $f$ and $g$ projection. In addition, there are $U(1)^6 \otimes U(1)^2$
gauge bosons and $12$ scalars in the ${\bf Z}_2$ orbifold.
The $g$ twist further projects out $4$ of the $U(1)^6$ gauge bosons and $8$ of
the scalars. Therefore, the ${\bf Z}_2 \otimes {\bf Z}_2^{\prime}$ orbifold
has $U(1)^2 \otimes U(1)^2$ gauge bosons and $4$ scalars.

{}The untwisted R-R sector ($V_1+V_2$ sector)
also gives rise to $U(1)$ gauge fields. They differ
from those in the NS-NS sector in that they are
obtained by tensoring the left- and right-moving spinor states.
As a result, there are $U(1)^8$ gauge bosons and $16$ scalars in the 
${\bf Z}_2$ orbifold.
The $g$ twist projects out half of the spectrum
and so there are $U(1)^4$ gauge bosons and $8$ scalars in the R-R
sector of the non-supersymmetric model.

{}The $f$ twist breaks the ${\cal N}=(4,4)$ supersymmetry to ${\cal N}=(2,0)$.
There are therefore $2$ gravitinos from the R-NS sector ($V_1$ sector),
but not in the NS-R sector ($V_2$ sector). In addition, there
are some spinors in both $V_1$ and $V_2$ sector.
Notice that the number of bosonic and fermionic degrees of freedom
cancel among the untwisted sector states even in the
non-supersymmetric ${\bf Z}_2 \otimes {\bf Z}_2^{\prime}$ orbifold.

{}There are also massless twisted sector states. In the ${\bf Z}_2$ orbifold
generated by $f$, the $f$ twisted NS-NS sector ($V_3$ sector)
provides $64$ scalars.
Their fermionic superpartners come from the $f$ twisted R-NS sector 
($V_3+V_1$) since
the left-moving supersymmetry is unbroken. Upon further projection by
the $g$ twist, there are $32$ scalars in the $V_3$ sector.
Notice that even the left-moving gravitinos are projected out by $g$,
the fermionic superpartners of the $32$ scalars still come from
$V_3+V_1$ sector.

{}There are other massless twisted sector states in the non-supersymmetric
model that are absent in the ${\bf Z}_2$ orbifold.
They are the massless states coming from the $fg$ twisted sector.
In Model IA, they come from the sectors 
$V_3+V_4 + \alpha_1 V_1 + \alpha_2 V_2$
for $\alpha_i = 0,1$. Again, there are equal number of bosons and fermions
among those sectors.
The massless spectrum of the model is summarized in Table \ref{IA}.
Here we emphasize that Model IA refers to more than one model:\\
$\bullet$ $(i)$ We leave the structure constants $k_{13}$ and $k_{24}$
unspecified. Different choices of $k_{13}$ and $k_{24}$ correspond to
different GSO projections, and hence the internal
quantum numbers of the states that are kept are different. 
The quantum numbers can be worked out easily by using the
spectrum generating formula in Ref.\cite{KLST}. 
However, the resulting spectra have the
same counting of the states, {\em i.e.}, the models have the
same number of scalars, spinors and $U(1)$ gauge bosons from each
sector.
We will see in Section \ref{Duality} that
models with different $k_{13}$ and $k_{24}$ have different duals. \\
$\bullet$ $(ii)$ The last $T^2$ are not touched by the ${\bf Z}_2$
twists $f$ and $g$. We can include shifts in this $T^2$
without affecting the supersymmetries of the individual ${\bf Z}_2$
orbifolds as well as the bose-fermi degeneracy feature of
the final ${\bf Z}_2 \otimes {\bf Z}_2^{\prime}$
orbifold. We will consider some variations by including shifts in
the following. The massless spectra for the models with shifts
are slightly different from the one without shift (some twisted
sector states become massive). Nonetheless, including shifts will not affect
our discussions of duality in Section \ref{Duality}.

{}Let us consider some variations of the above model.
In the above, we have assumed that the worldsheet
fermions can always be written in the complex basis.
This is not necessary for worldsheet consistencies if only
${\bf Z}_2$ twists are involved. As mentioned before
in Section \ref{prelim}, the complex fermions can be split into
pairs of real fermions with different boundary conditions, as long
as the {\em cubic} constraint (Eq.~(\ref{cubic}))
is satisifed. This opens up the possibilities for other models.
Consider a variation of Model IA with the following generating vector:
\begin{eqnarray}
V_3 &=& \textstyle{\left[ 0 ~( -{1\over 2} -{1\over 2} ~0 )^2
                               (0 0 0)_r (0 -{1\over 2} -{1\over 2})_r 
      \vert \vert  0 ~(0 -{1\over 2} -{1\over 2})^2 (0 0 0)_r 
                          (0 -{1\over 2} -{1\over 2})_r \right]}~, \\
V_4 &=& \textstyle{\left[ 0 ~(0 -{1\over 2} -{1\over 2})^2 
  (0 -{1\over 2} -{1\over 2})_r (0 0 0)_r 
\vert \vert
0 ~( -{1\over 2} -{1\over 2} ~0 )^2
  (0 -{1\over 2} -{1\over 2})_r  (0 0 0)_r \right]} ~. 
\end{eqnarray}
where the subscript $r$ indicates that the corresponding worldsheet
fermions are real, the worldsheet fermions are complex otherwise. 
Clearly, the cubic constraint is satisfied. This model
is related to the model 
presented in Ref.~\cite{KKS}. This can be seen as follows. Since we
have fermionized a real boson $\phi$ into
a pair of real fermions $\psi_1$ and $\psi_2$:
\begin{equation}
\psi_1 + i \psi_2 = e^{i \phi}
\end{equation}
The ${\bf Z}_2$ twist on the fermions: 
$\psi_1 \rightarrow - \psi_1$
and $\psi_2 \rightarrow - \psi_2$ corresponds to a
${\bf Z}_2$ {\em shift} on the boson: 
$\phi \rightarrow \phi + \pi$.
The ${\bf Z}_2$ twist on the fermions: 
$\psi_1 \rightarrow \psi_1$ and $\psi_2 \rightarrow
- \psi_2$ corresponds to a ${\bf Z}_2$ {\em twist} on the boson:
$\phi \rightarrow - \phi$. In the orbifold langauge, the $f$ and $g$ twists
can be written as:
\begin{eqnarray}
f &=& \left[ (-1^4,1^2),(1^6),(0^4,0,s),(s^4,0,s), (-1)^{F_R} \right]~, \\
g &=& \left[ (1^6),(-1^4,1^2),(s^4,s,0),(0^4,s,0), (-1)^{F_L} \right]~.
\end{eqnarray}
Here, the element of the space group of the orbifold is
denoted by
\begin{equation}
\left[ (\theta_L),(\theta_R),(v_L),(v_R),\Theta \right]
\end{equation}
where $\theta_{L,R}$ are rotations by elements of $SO(6)$, and
$v_{L,R}$ are shifts acting on 
the left- and right-moving degrees of freedom respectively. The eigenvalue
of $\Theta$ can be $1$ or $(-1)^{F_{L,R}}$ depending on whether
there is discrete torsion.

{}Notice that the variation (corresponding to a shift on the last circle)
does not affect the supersymmetries of the individual ${\bf Z}_2$ orbifolds
generated by $f$, $g$ and $fg$, as well as the final non-supersymmetric
${\bf Z}_2 \otimes {\bf Z}_2^{\prime}$ orbifold (since the extra
shift has no effect on the gravitinos). Futhermore, the requirement
for vanishing one-loop cosmological constant (Eq.~(\ref{cross})) 
is preserved. The multi-loop analysis (see Section \ref{multiloop})
also remains unchanged by the
modification.

{}The model has the same massless untwisted sector spectrum as
Model IA. However, some of the twisted sector states become massive
because of the extra shift.
The massless twisted sector states come only from the following
$4$ sectors: $V_3 + V_4 + V_0 + \sum_{i=1}^2 \alpha_i V_i$ where 
$\alpha_i=0,1$.
The $V_3+V_4+V_0$ sector provides $U(1)^8$ gauge bosons and
$16$ real scalars, $V_3+V_4+V_0 + V_1 + V_2$ provides $32$ scalars.
The sectors $V_3 + V_4 + V_0+ V_1$ and $V_3+V_4+V_0+V_2$
each provides $16$ spinors.
The number of bosonic and fermionic degrees of freedom are equal in
the twisted sectors.

{}We can construct yet another variation of Model IA with the following
generating vectors:
\begin{eqnarray}
V_3 &=& \textstyle{\left[ 0 ~( -{1\over 2} -{1\over 2} ~0 )^2
                               (0 0 0)_r (0 -{1\over 2} -{1\over 2})_r 
      \vert \vert  0 ~(0 -{1\over 2} -{1\over 2})^2 (0 0 0)_r 
                          (0 -{1\over 2} -{1\over 2})_r \right]}~, \\
V_4 &=& \textstyle{\left[ 0 ~(0 -{1\over 2} -{1\over 2})^2 (0 0 0)_r 
  (0 -{1\over 2} -{1\over 2})_r 
\vert \vert
0 ~( -{1\over 2} -{1\over 2} ~0 )^2 (0 0 0)_r
  (0 -{1\over 2} -{1\over 2})_r  \right]} ~. 
\end{eqnarray}
which in the orbifold language corresponds to:
\begin{eqnarray}
f &=& \left[ (-1^4,1^2),(1^6),(0^4,0,s),(s^4,0,s), (-1)^{F_R} \right]~, \\
g &=& \left[ (1^6),(-1^4,1^2),(s^4,0,s),(0^4,0,s), (-1)^{F_L} \right]~.
\end{eqnarray}
A similar model has been considered in Ref.~\cite{harvey}.
Again, it has the same untwisted sector massless spectrum. The twisted 
sector massless states come from the sectors
$V_3+ V_4 + \sum_{i=1}^2 \alpha_i V_i$. The number of bosonic
and fermionic degrees of freedom are equal in the twisted 
sectors.

{}In Model IA and two of its variations above, the radii of the last $T^2$ 
are not fixed by the asymmetric orbifold.
Changing the radii does not change the bose-fermi degeneracy feature of the
resulting ${\bf Z}_2 \otimes {\bf Z}_2^{\prime}$ models, and hence 
they are
free moduli.
Starting from the model constructed above by the free fermionic approach,
which fixes the radii to be 1 \footnote{We use the convention
that $P_{L,R}={m \over 2R} \pm nR$ for $m,n \in {\bf Z}$.},
we can freely change the complex and the K\"ahler moduli of the
last $T^2$ while maintaining the bose-fermi degeneracy.

{}Clearly, many more variations of the above model with the bose-fermi
degenerate feature can be constructed
using real worldsheet fermions. It would be interesting to
work out the models in detail, and perhaps classfiy all model with
this feature. \\

$\bullet$ \underline{Model IB} 
\begin{eqnarray}
V_3 &=& \textstyle{\left[ 0 ~( -{1\over 2} -{1\over 2} ~0 )^2
                               (0 0 0) 
      \vert \vert  0 ~(0 -{1\over 2} -{1\over 2})^2 (0 0 0) \right]}~, \\
V_4 &=& \textstyle{\left[ 0 ~(0 0 0) (0 -{1\over 2} -{1\over 2})^2  
\vert \vert
0 ~(0 0 0)( -{1\over 2} -{1\over 2} ~0 )^2
                                \right]} ~. 
\end{eqnarray}

{}The massless spectrum of the model is given in Table \ref{IB}.
It is identical to that of Model IA except that
the $fg$ twisted sector states come from the
$V_3+V_4 +V_0 + \alpha_1 V_1 + \alpha_2 V_2$ sectors.
Similarly, one can construct variations of the above model
by using real worldsheet fermions.
An interesting feature of this model 
is that all the radii of $T^6$
are fixed by the asymmetric orbifold (as compared with Model IA
in which the radii of the last two component of $T^6$ are free) and
yet gives rise to the same counting of the massless spectrum as Model IA.

\subsection{Class II}


{}The general forms of $V_3$ and $V_4$ are 
given by:
\begin{eqnarray}
V_3 &=& \textstyle{\left[ 0 \left( -{1\over 2} -{1\over 2} ~0 \right)^2
                               (0 ~d_{3} ~d_{3}) 
      \vert \vert  0 ~(0 ~a_3 ~a_3 ) (0 ~b_3 ~b_3) (0 ~c_3 ~c_3) 
\right]}~, \\
V_4 &=& \textstyle{\left[ 0 ~(0 ~a_4 ~a_4 ) (0 ~b_4 ~b_4) (0 ~c_4 ~c_4) 
\vert \vert
 0 ~(0 ~a^{\prime}_4 ~a^{\prime}_4 ) (0 ~b_4^{\prime} ~b^{\prime}_4) 
(0 ~c^{\prime}_4 ~c^{\prime}_4)
                             \right]} ~. 
\end{eqnarray}
The structure constants $k_{23}=1/2$, $k_{14}=1/2$, $k_{24}=0$, whereas
the other $k_{ij}$ are not fixed by the supersymmetries of the
individual ${\bf Z}_2$ orbifolds.
The monodromies must be chosen such
that $V_i^2 \in {\bf Z}$,
$V_3 \cdot V_4 = 0~\mbox{or}~1/2~(\mbox{mod}~1)$ and
Eq.~(\ref{cross}) is satisfied. There are three models in this
class. The massless spectra of these models are given in the tables.
Here, we only list
the generating vectors corresponding to the twists $f$ and $g$: \\

$\bullet$ \underline{Model IIA} 
\begin{eqnarray}
V_3 &=& \textstyle{\left[ 0 ~( -{1\over 2} -{1\over 2} ~0 )^2
                               (0 0 0) 
      \vert \vert  0 ~(0 -{1\over 2} -{1\over 2})^2 (0 0 0) \right]}~, \\
V_4 &=& \textstyle{\left[ 0 ~(0 -{1\over 2} -{1\over 2})^2 (000)
\vert \vert
0 ~(0 0 0)^3                                \right]} ~. 
\end{eqnarray}

The massless spectrum of this model is given in Table \ref{IIA}.\\

$\bullet$ \underline{Model IIB} 
\begin{eqnarray}
V_3 &=& \textstyle{\left[ 0 ~( -{1\over 2} -{1\over 2} ~0 )^2
                               (0 0 0) 
      \vert \vert  0 ~(0 -{1\over 2} -{1\over 2})^2 (0 0 0) \right]}~, \\
V_4 &=& \textstyle{\left[ 0 ~(0 00)^3
\vert \vert
0 ~(0 0 0) (0 -{1\over 2} -{1\over 2})^2                 \right]} ~. 
\end{eqnarray}

The massless spectrum of this model is given in Table \ref{IIB}.\\

$\bullet$ \underline{Model IIC} 
\begin{eqnarray}
V_3 &=& \textstyle{\left[ 0 ~( -{1\over 2} -{1\over 2} ~0 )^2
                               (0 0 0) 
      \vert \vert  0 ~(0 -{1\over 2} -{1\over 2})^2 (0 0 0) \right]}~, \\
V_4 &=& \textstyle{\left[ 0 ~(0 00)^2 ( 0 -{1\over 2} -{1\over 2})
\vert \vert
0 ~(0 -{1\over 2} -{1\over 2}) (0 0 0)^2                 \right]} ~. 
\end{eqnarray}

The massless spectrum of this model is given in Table \ref{IIC}. \\

{}Similar to the set of models in Class I, one can construct variations of 
the above models by using real worldsheet fermions. Notice that the
structure constant $k_{13}$ is unspecified, different choices of $k_{13}$
correspond to different GSO projection, and hence different spectra.
However, the number of scalars, spinors and $U(1)$ gauge bosons 
from each sector are not altered.

{}If the $N$-loop vacuum amplitude vanishes, the one-, the two-, and the 
three-point $N$-loop amplitudes generically vanish as well. 
In particular, since 
the mass corrections come from the two-point $N$-loop amplitudes, the 
vanishing $N$-loop vacuum amplitude implies vanishing mass corrections.
If the cosmological constant vanishes perturbatively,
the tree level mass spectrum is not corrected perturbatively.

{}There are massless twisted sector states in this set of models
(in both Class I and II).
Despite the fact that the model is non-supersymmetric (since the
gravitinos are projected out), the twisted sector states seem to fall into
appropriate supermultiplets (depending on the number of supersymmetries
that survive the orbifold twist of the sector under consideration).
Whether the twisted sector states indeed assemble themselves into
supermultiplets
can be confirmed with an explicit calculation of the couplings
between the various massless states.
This may not be surprising since
the gravitinos are projected out by the discrete torsions $(-1)^{F_L}$
(associated with the $g$ twist) and $(-1)^{F_R}$ (associated with the
$f$ twist). If at least one of the discrete torsions is absent, then
the model is supersymmetric.
Each ${\bf Z}_2$ twist projects out half of the spectrum.
The discrete torsion interchanges the role of the
half which is projected out and the other half which is kept.
The number of bosons as well as the number of fermions
in each half of the spectrum are nonetheless
the same. Hence the bose-fermi degeneracy in the non-supersymmetric
model has a supersymmetric origin.
Even though the $4$-dimensional gravitinos $\chi^{\mu}_a$ are all
projected out with the particular choice of discrete torsion,
the same number of fermionic degrees of freedom is restored by
the spinors $\chi^i_a$ that are originally
projected out in the supersymmetric model.
Upon decompactification of $T^6$, these fermions
become part of the $10$-dimensional
gravitinos.

{}As pointed out in Ref\cite{KKS,KS} and the discussions in the next 
sections here, at least some of these models are expected to have exactly 
zero cosmological constants. 
The vanishing of the cosmological constant in these
models has a natural explanation in string theory--- it is simply
a remnant of the $10$-dimensional supersymmetry that defines
Type II string theory. In $4$-dimensional supersymmetric string models, 
the remnant of the $10$-dimensional supersymmetry lives in the spacetime 
sector, yielding lower spacetime sypersymmetry. Here, the remnant of the 
$10$-dimensional supersymmetry somehow lives inside the internal space, 
yielding non-supersymmetric models.
The exact form of this remnant symmetry remains to be understood.

\section{Duality and Self-Duality}\label{Duality}

{}Finally we are ready to address the issue of duality of Model IA.
As emphasized before, Model IA refers to more than one model
since some of the discrete torsions which do not affect the counting of
the spectrum (corresponding to the
values of $k_{13}$ and $k_{24}$) are unspecified.
However, these discrete torsions turn out to be
crucial when we address the issue
of duality because models with different discrete torsions
have different duals.
In particular, there are three independent choices:
$(i)$ Model IA$_i$: $k_{13}=0$, $k_{24}=1/2$;
$(ii)$ Model IA$_{ii}$: $k_{13}=1/2$, $k_{24}=0$; and
$(iii)$ Model IA$_{iii}$: $k_{13}=k_{24}=0$.
(The remaining choice
$k_{13}=k_{24}=1/2$ is related to Model IA$_{iii}$ by a reflection
of left- and right-moving quantum numbers).
Before we go into the discussion, let us summarize the final conclusion:
Model IA$_{i}$ and IA$_{ii}$ are both 
self-dual (although the $f$ and $g$ twisted
sectors are interchanged by duality in Model IA$_{ii}$).
Model IA$_{iii}$ and Model IIA are a strong-weak dual ({\em i.e.}, $U$-dual)
pair. 
Let us first briefly review the relevant 
ingredients of the construction of Type II string dual pairs 
\cite{SV} and the particular application used in Ref\cite{KS}.
We follow closely their notation.

{}A systematic construction of dual pairs of type II compactifications in
$D=4$ dimensions was discussed by Sen and Vafa \cite{SV}.
The U-duality group of Type II compactifications on $T^4$ is
$SO(5,5; {\bf Z})$, while the perturbatively obvious T-duality subgroup
is $SO(4,4;{\bf Z})$.
Consider two elements $h,\tilde{h} \in SO(4,4;{\bf Z})$,
which are not conjugate in $SO(4,4;{\bf Z})$ but which are conjugate
in $SO(5,5;{\bf Z})$:
\begin{equation}\label{congugate}
g h g^{-1} = \tilde{h},~~g \in SO(5,5;{\bf Z})
\end{equation}
The particular $SO(5,5;{\bf Z})$ element $\overline{g}$ of interest is
given in terms of the element $\sigma$ of the ten-dimensional 
$SL(2,{\bf Z})$
symmetry group of Type IIB which inverts the ten-dimensional
axion/dilaton field and the T-duality element $\tau_{1234}$ that inverts
the volume of the $T^4$:
\begin{equation}\label{barg}
\overline{g} ~=~\sigma \cdot \tau_{1234}\cdot \sigma^{-1}
\end{equation}
This maps fundamental strings (without winding on the $T^4$) to
NS fivebranes wrapped on the $T^4$.
The element $\overline{g}$ has the property that
\begin{equation}\label{helpful}
\overline{g} h {\overline{g}}^{-1}  \in SO(4,4;{\bf Z})
\end{equation}
for all $h \in SO(4,4;{\bf Z})$ \cite{SV}.

{}Now, consider
compactifying on an additional $T^2$, which we can
take to be a product of two circles.  
We can orbifold the resulting
compactifications by $h$ ($\tilde{h}$) acting on the $T^4$.
By the adiabatic argument \cite{VW}, the resulting models in
four dimensions are still dual.
In fact, the dual models that the adiabatic argument yields will be
related by $S-T$ exchange:
\begin{equation}\label{ST}
\tilde{S} ~=~ T,~~\tilde{T} ~=~ S
\end{equation}
where $T$ is the K\"ahler modulus associated with the $T^2$ and $S$ is the
axion-dilaton in four dimensions. 

{}Consider an element $h$ of $SO(4,4;{\bf Z})$ which acts on
$X^{1...4}_L, X^{1...4}_R$ as pairwise rotation, labeled by the four angles
$(\theta_L,\phi_L,\theta_R,\phi_R)$.
Then ${\overline{g}}$ conjugates $h$ to $\tilde{h}$ which acts on
$X^{1...4}_L, X^{1...4}_R$ as $(\tilde \theta_L, \tilde \phi_L,
\tilde \theta_R, \tilde \phi_R)$
where
\begin{equation}\label{dualmatrix}
\pmatrix{\tilde{\theta_L} \cr
                   \tilde{\phi}_L \cr
                   \tilde{\theta}_R \cr
                   \tilde{\phi}_R}
~=~ \pmatrix{1/2 & -1/2 & 1/2 & -1/2 \cr
             -1/2 & 1/2 & 1/2 & -1/2 \cr
             1/2 & 1/2 & 1/2 & 1/2 \cr
             -1/2 & -1/2 & 1/2 & 1/2}
~\pmatrix{\theta_L \cr
          \phi_L \cr
          \theta_R \cr
          \phi_R}
\end{equation}
is the equation that will yield the Type II duals of our orbifolds.

{}Now let us concentrate on the action on the first $T^4$. In the above 
notation, $V_3$ in Model IA (or the KKS model) 
can be represented as
\begin{equation}\label{f}
f ~=~ (\pi, \pm \pi,2\pi,0),
\end{equation}
where the $\pi$ corresponds to a $Z_2$ twist, $(-1)^{F_R}$ 
({\em i.e.}, the discrete torsion from the structure constant $k_{23}=1/2$) 
is represented by a $2\pi$ rotation on right movers, and the choice
$-\pi$ corresponds to a ${\bf Z}_2$ twist with a discrete torsion $(-1)^{F_L}$
({\em i.e.}, the discrete torsion from the structure constant $k_{13}=1/2$).
The $0$ implies no twist but still there can be shifts.
In both cases in Eq.~(\ref{f}), the left-moving 
supersymmetry is broken to ${\cal N}=2$ (the discrete torsion determines
which $2$ of the gravitinos are kept). Therefore, in the previous 
section, we leave it unspecified.
However, 
these two cases transform differently under the 
duality transformation Eq.~(\ref{dualmatrix}), which we will now
discuss.

{}For $f=(\pi,\pi,2 \pi,0)$, from the action of Eq.~(\ref{dualmatrix}),
we see that $\tilde{f} = f$, implying that the $(2,0)$ model 
(${\cal N}_L=2$ and ${\cal N}_R=0$) is self-dual. 
On the other hand, if $f=(\pi,-\pi,2\pi,0)$, then
$\tilde{f}=(2\pi,0,\pi,\pi)$. Therefore, the $(2,0)$ model is mapped
to the $(0,2)$ model.

{}Similarly, one can consider the duality transformation of the other
${\bf Z}_2$ twists $g$ and
$fg$. 
It turns out there are three independent cases
for the resulting ${\bf Z}_2 \otimes {\bf Z}_2^{\prime}$ orbifold:\\

\noindent $\bullet$ $(i)$ $f = (\pi,\pi,2\pi,0)$, $g = (2\pi,0,\pi,-\pi)$
and $fg=(-\pi,\pi,-\pi,-\pi)$. 

\begin{center}
\hspace{2cm}
\epsfxsize=16 cm
\epsfbox{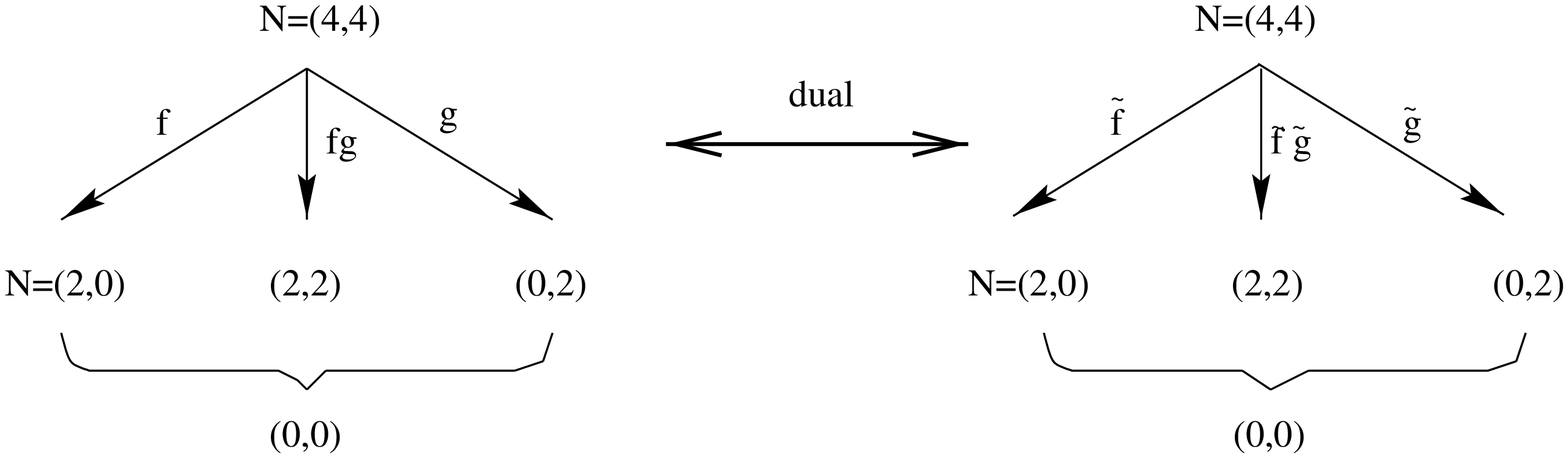}
\end{center}

{}Duality transformation Eq.~(\ref{dualmatrix})
maps them to $\tilde{f}=f$, $\tilde{g}=g$ and $\tilde{f} \tilde{g}= fg$.
The ${\bf Z}_2 \otimes {\bf Z}_2$ orbifold is self-dual.
This is the dual pair (KKS Model) considered in Ref. \cite{KS}. 
With the appropriate choices of $k_{ij}$ ({\em i.e.}, 
$k_{13}=0$ and $k_{24}=1/2$), 
Model IA also belongs to this type. Let us denote the corresponding 
model
by Model IA$_{i}$.\\

\noindent $\bullet$ $(ii)$ $f=(\pi,-\pi,2\pi,0)$, $g=(2 \pi,0,\pi,\pi)$
and $fg=(-\pi,-\pi,-\pi,\pi)$. 

\begin{center}
\hspace{2cm}
\epsfxsize=16 cm
\epsfbox{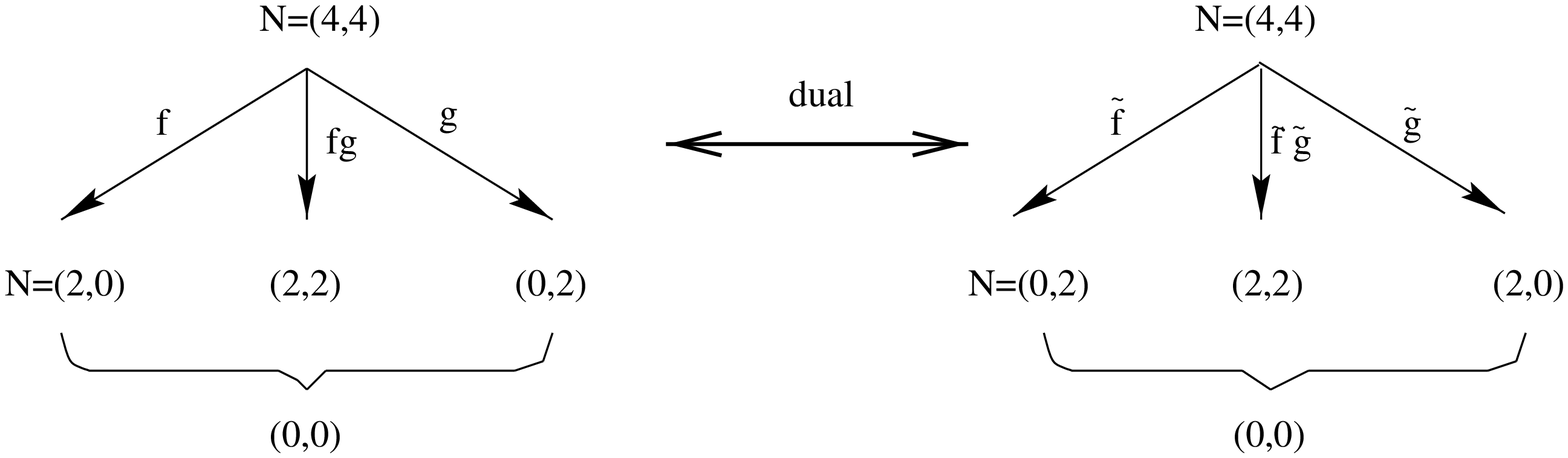}
\end{center}

{}Duality transformation maps
them to $\tilde{f}=g$, $\tilde{g}=f$ and
$\tilde{f} \tilde{g}=fg$. This corresponds to $k_{13}=1/2$ and
$k_{24}=0$. Let us denote the corresponding 
${\bf Z}_2 \otimes {\bf Z}_2^{\prime}$ orbifold by Model IA$_{ii}$.
The model
is still self-dual, but the $f$ and $g$ twist sectors are interchanged. \\

\noindent $\bullet$ $(iii)$ $f=(\pi,\pi,2\pi,0)$, $g=(2\pi,0,\pi,\pi)$
and $fg=(-\pi,\pi,-\pi,\pi)$. 

\begin{center}
\hspace{2cm}
\epsfxsize=16 cm
\epsfbox{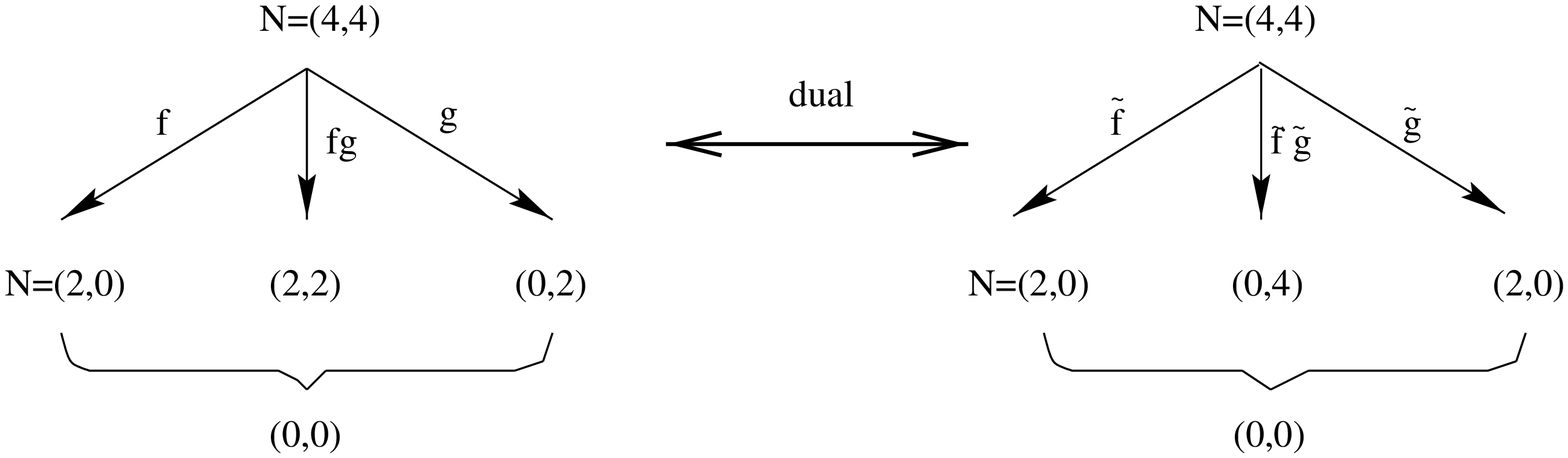}
\end{center}

{}Duality transformation maps them
to $\tilde{f}=f$, $\tilde{g}=(\pi,-\pi,2\pi,0)$
and $\tilde{f}\tilde{g}=(2\pi,0,0,0)$. 
This corresponds to $k_{13}=k_{24}=0$. We denote the
corresponding ${\bf Z}_2 \otimes {\bf Z}_2^{\prime}$ orbifold
by Model IA$_{iii}$.
In this case, the dual pair is more non-trivial. 
In the dual model, the twists $\tilde{f}$, $\tilde{g}$ and 
$\tilde{f} \tilde{g}$ break supersymmetry to ${\cal N}=(2,0)$,
${\cal N}=(2,0)$ and ${\cal N}=(0,4)$ respectively.
Since duality maps the $(2,2)$ model to the $(0,4)$ model, the $U(1)$
gauge fields from the NS-NS sector in one model would 
be mapped to those in the R-R sector of its dual.
 
{}Notice that even though the duality transformation can map 
the individual ${\bf Z}_2$ orbifolds to orbifolds with {\em different}
supersymmetries, the resulting model ${\bf Z}_2 \otimes {\bf Z}_2^{\prime}$
orbifold is still non-supersymmetric. This can be seen from
Eq.~(\ref{sum}). In particular, consider the contribution of the
gravitinos in the partition function:
\begin{eqnarray}
({\cal N}_L,{\cal N}_R) &=& {1\over 2}~ (2,0) + {1\over 2}~(0,2) + {1\over 2}
~(2,2) - {1\over 2} ~(4,4) = (0,0) \nonumber \\
&\rightarrow& {1\over 2}~ (2,0) + {1\over 2}~(0,2) + {1\over 2}
~(2,2) - {1\over 2} ~(4,4) = (0,0) \nonumber  \\
&\rightarrow& {1\over 2}~ (0,2) + {1\over 2}~(2,0) + {1\over 2}
~(2,2) - {1\over 2} ~(4,4) = (0,0) \nonumber  \\
&\rightarrow& {1\over 2}~ (2,0) + {1\over 2}~(2,0) + {1\over 2}
~(0,4) - {1\over 2} ~(4,4) = (0,0) \nonumber
\end{eqnarray}

{}Therefore, the dual model is also non-supersymmetric.
Furthermore, if the dual model satisfies also the requirement
Eq.~(\ref{cross}), then the model and its dual both have 
vanishing one-loop cosmological constant.
This requirement is trivially satisfied for cases $(i)$ and $(ii)$,
but {\em a priori}, it may not be satisfied in case $(iii)$.
However, a closer examination of the structure clearly shows that the dual 
model also has bose-fermi degeneracy. 

{}To find its dual, we first notice that
the $T^2$ in Model IA is untouched, meaning that it is toroidally 
compactified. To find its dual, we need to have appropriate twists such 
that the same $T^2$ is untouched. 
A natural candidate for the dual of Model IA$_{iii}$ with the above
mentioned properties is Mode IIA. In Model IIA, the generating 
vectors $V_3$, $V_4$ and $V_3+V_4$ (corresponding to three 
different ${\bf Z}_2$ twists) break supersymmetry to ${\cal N}=(2,0)$,$(0,4)$ 
and $(2,0)$ respectively. We can identify $\tilde{f}$ with $V_3$,
$\tilde{g}$ with $V_3+V_4$ and $\tilde{f} \tilde{g}$ with
$V_4$. As we have shown in the previous section, the 
requirement Eq.~(\ref{cross}) is satisifed in Model IIA and so it
has vanishing
one-loop cosmological constant.
Thus, the dual of Model IA$_{iii}$ is a {\em different} orbifold
but yet exhibits the same perturbative vanishing of the cosmological
constant as the the original
model.
The massless spectrum of Model IIA is given in
Table \ref{IIA}. Despite the fact that Model IA$_{iii}$ and Model IIA
have the same counting of states, a careful examination of the
spectrum (using the spectrum generating formula in Ref.\cite{KLST})
indicates that they have different internal quantum numbers, and hence
different spectra. 

{}Let us summarize the duality relations between different models
as follows: \\

\begin{center}
\hspace{1.5cm}
\epsfxsize=10 cm
\epsfbox{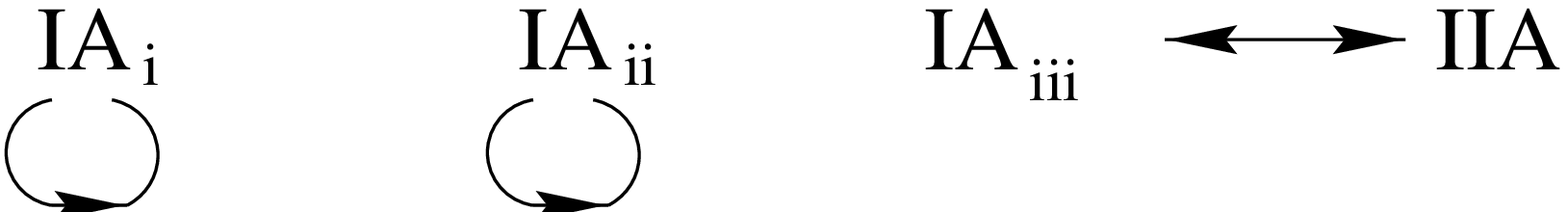}
\end{center}

{}In both Model IA and Model IIA, 
the radii of the last $T^2$ are not fixed by the 
asymmetric orbifold.
Changing the radii of the
last $T^2$ does not affect the supersymmetries of the individual ${\bf Z}_2$
orbifolds as well as the bose-fermi degeneracy feature of the resulting
${\bf Z}_2 \otimes {\bf Z}_2^{\prime}$ orbifold and hence
they are free moduli. 
The string coupling $S$ and the K\"ahler modulus associated with the
$T^2$, {\em i.e.}, $T$, are exchanged
in the dual model by Eq.~(\ref{ST}).
This allows us to argue that in various strongly coupled limits of
the original model, there exists weakly coupled duals which exhibit the same
perturbative cancellations as the original model:\\
$\bullet$ For $S \rightarrow \infty$ and arbitrary $T$, 
the orignial model is weakly coupled.\\
$\bullet$ For $T \rightarrow \infty$ and arbitrary $S$, one can go to 
the dual with $\tilde{S}=T \rightarrow \infty$ and $\tilde{T}=S$ which
is weakly coupled and perturbative contributions cancel in a similar
way as the origninal model.\\
$\bullet$ For $S \rightarrow 0$ and $T \rightarrow 0$, the original model
is strongly coupled and at a small radius. One can $T$-dualize $T^2$ to
get a model at strong coupling but with large radius $T \rightarrow \infty$.
From the $S-T$ exchange (\ref{ST}), 
the dual model is weakly coupled 
$\tilde{S} \rightarrow \infty$ and at small radius.

{}One can further deduce from string duality
that at least some of the non-perturbative
corrections to the cosmological constant are absent.
{\em A priori},
there can be non-perturbative corrections to the cosmological constant
for finite $S$.
For instance,
there may be non-zero contribution due to an unequal number (or masses) of
bosonic and fermionic solitonic
states coming from the NS fivebranes wrapped
on $T^4$.
Let us consider the limit when $T$ is large.
In this limit, the dual model is weakly coupled
($\tilde{S}=T \rightarrow \infty$)
and the perturbative 
contributions in the dual model vanish for arbitrary
$\tilde{T}=S$ due to the bose-fermi degeneracy, while the non-perturbative
contributions in the dual model are exponentially suppressed.
Since Eq.~(\ref{barg}) maps the NS fivebranes wrapped on $T^4$ to the
fundamental strings (without winding on the $T^4$)
in the dual model \cite{SV}, the 
contribution from this type of solitonic states also obey bose-fermi
degeneracy. As a result, the
non-perturbative corrections due to this type of wrapped NS fivebranes 
are absent due to bose-fermi degeneracy. 

 {}It will be interesting to find out whether the IA model has exactly zero 
cosmological constant or not, that is, whether all its solitionic 
contributions obey bose-fermi degeneracy. As first pointed out by Harvey 
\cite{harvey}, duality is a powerful tool in addressing this issue.
However, it is not clear that all duals are relevant to the above question.
In particular, the heterotic dual \cite{harvey} of a Type II model with 
bose-fermi degeneracy has a non-zero one-loop cosmological constant. 
Naively, this implies a non-vanishing dilaton potential, which is 
non-perturbative in the original Type II coupling. However, there is a 
barrier (hence a tachyon) as the dilaton expectation value interpolates 
between this particular dual pair. It is then not clear that, both away and 
around the barrier, the non-vanishing cosmological constant can be 
interpreted as coming from a set of solitonic states of the Type II model. 
If the dual pair is not connected by varying parameters in the 
non-perturbative string model, it may not shed light on the 
solitonic bose-fermi degeneracy issue.

{}A non-zero cosmological constant coming from the solitonic sector is 
exponentially suppressed in the coupling. Since the coupling in nature is 
rather small, a non-perturbative cosmological constant may be 
perfectly harmless phenomenologically. 

{}Since Model IA$_i$, IA$_{ii}$ and IA$_{iii}$ have different duality
properties under the map Eq.~(\ref{barg}), it is interesting to
understand further the relations between these models. Recall that these
models differ only in the structure constants $k_{13}$ and $k_{24}$
which can take values $0$ or $1/2$.
In the orbifold language, $k_{13}=1/2$ correponds to including 
$(-1)^{F_L}$ in the action of $f$. Similarly, $k_{24}=1/2$ corresponds to
including $(-1)^{F_R}$ in the action of $g$. 
Let us start with Model IA$_{iii}$ which has $k_{13}=k_{24}=0$.
Consider $T$-dualizing one
of the circles in $T^2$, {\em i.e.} $R \rightarrow 1/R$. Since $T$-duality
is a one-sided parity operation, the action on the left- and right-moving
coordinates is given by $X_L \rightarrow X_L$ and $X_R \rightarrow -X_R$.
The corresponding action on the worldsheet
fermions (from worldsheet supersymmetry) is $(-1)^{F_R}$ (since it 
anti-commutes with $\psi_R$). In other words, Model IA$_{i}$ (with 
$k_{13}=0$ and $k_{24}=1/2$) can be
obtained from Model IA$_{iii}$ by $T$-dualizing the action of $g$.
Similarly, Model IA$_{ii}$ can be obtained from Model IA$_{iii}$
by first interchanging the left- and right-moving worldsheet 
degrees of freedom
and then $T$-dualizing the action of
$f$. As a result, the various models are then 
related by 
an intricate web of dualities: 

\begin{center}
\hspace{1.5cm}
\epsfxsize=10 cm
\epsfbox{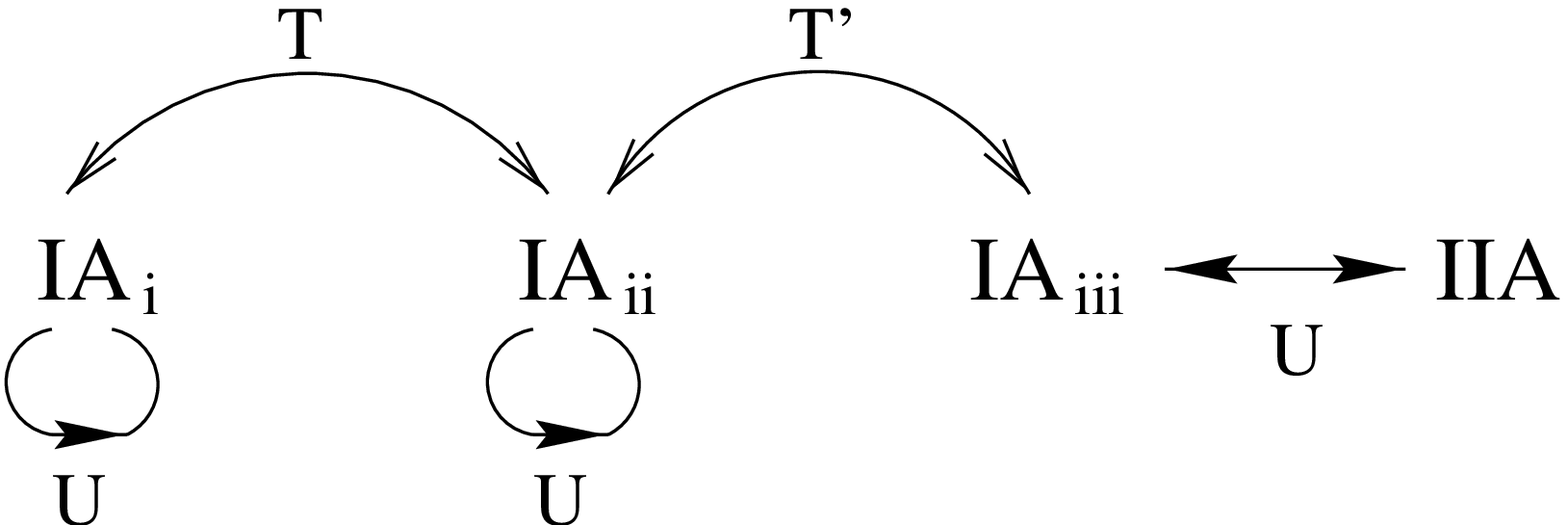}
\end{center}
where $U$ is the $U$-duality operator defined in Eq.~(\ref{barg}),
$T$ and $T^{\prime}$ are two different $T$-duality operators.

{}It would be interesting to work out the duality properties of the other
models ({\em i.e.} IB, IIB and IIC) presented in this paper.

\section{Multi-Loop Amplitudes}\label{multiloop}

{}The question whether the cosmological constants of the models constructed 
above vanish beyond one-loop is clearly important. Unfortunately, we are 
not able to give a definitive answer to this question, due to the subtleties 
involved in the 
multi-loop calculations. We can summarize the situation as follows. The 
calculation of the two-loop cosmological constant of the above models is 
exactly 
parallel to that of the model in Ref\cite{KKS}. If we 
take the gauge choice in Ref. \cite{KKS}, it is straightforward 
to demonstrate the vanishing of the cosmological constant; 
in fact, the two-loop integrand vanishes point-wise in the moduli space. 
A naive generalization of the two-loop argument actually yields the 
point-wise vanishing of the multi-loop integrand in the vacuum amplitude. 
However, there is no proof of the validity of the particular gauge
choice beyond two-loops. 
So, whether the cosmological constants of the these models vanish beyond 
one-loop hinges on the legitimacy of the particular gauge choice. 
Since we do not have any particularly intelligent observations 
to make on this point, our discussions here will be brief.

{}In evaluating the g-loop string vacuum amplitude, we are evaluating the 
amplitude associated with genus-g Riemann surfaces. The amplitude has a 
holomorphic and a similar anti-holomorphic part, and we may consider them 
separately in the discussion. In the covariant 
formulation, it is necessary to insert $2(g-1)$ picture-changing operators, 
which correspond to the integration of the odd moduli and/or absorb the 
zero modes of the $\beta$ superghost \cite{VV}. Moving the 
positions $z_i$ of these insertion points changes the integrand in the vacuum 
amplitude by a total derivative. In this sense, we see the choice of $z_i$ 
as pure gauge choices. 
For arbitrary $z_i$, the modes along the longitudinal and 
the time directions are not cancelled by the (super-)ghost modes (except in 
the one-loop case, where there is no picture-changing operator insertion), 
{\em i.e.}, their respective theta functions do not cancel. 
However, in the light-cone gauge, we have neither longitudinal/time nor 
ghost modes. So we expect that there must exist a choice of the 
positions $z_i$ so that the theta functions from the longitudinal/time 
modes cancel that from the ghost modes. The choice that is closest to this 
picture is the so called "unitary" gauge \cite{AMS,LP,LP2},
\begin{equation}
\sum_{i=1}^{2(g-1)} z_i = 2 \Delta 
\end{equation} 
where $\Delta$ is a Riemannn constant. For 
generic values of $z_i$ satifying this unitary gauge condition, the theta 
functions still do not exactly cancel. In the two-loop case,
there is a $z_1 -z_2$ dependence in the integrand. 
Now let us look at the situation in the 
light-cone gauge. Recall that a picture-changing supercurrent must 
be inserted at each interaction vertex \cite{mandel}. Since there are 
$2(g-1)$ vertices in the g-loop diagram, we see that such insertions are an 
intrinsic part of the multi-loop light-cone amplitudes, {\em i.e.},
the $z_1-z_2$ dependence in the two-loop case is not unexpected.

{}The consistency of the unitary gauge has been demonstrated in 
Ref\cite{LP,LP2}. In particular, modular invariance has been shown for 
generic choices of $z_i$ satisfying the unitary gauge condition.
In Ref\cite{KKS}, the further choice is made for the two-loop case 
\cite{AMS,LP2}  
\begin{equation}
z_1 - z_2 = 0
\end{equation}
This choice is perfectly reasonable. Since the cosmological 
constant does not depend on $z_1 -z_2$, and there is no integration over 
$z_i$ in the evaluation of the cosmological constant, the 
$z_1-z_2$ dependence must be a pure gauge choice. Different choices simply
gives different total derivative terms, and the total derivative term in 
this gauge is simply zero. Now, under the Dehn twist, the phase factor
$2\pi i\left[\alpha(s+{\overline s}) +\gamma(s+{\overline s})\right]$
in Eq.~(\ref{mod2}) needed for modular invariance is apparently absent in 
this $z_1=z_2$ gauge, {\em i.e.}, modular invariance is obscure. 
To resolve this ambiguity, we shall employ the following point-split 
approach. To carry out the Dehn twist, we first split $z_1$ and $z_2$, do 
the modular transformation, and then take the limit $z_1=z_2$ again.
In this gauge choice, the cosmological constant 
of the non-supersymmetric string model constructed in Ref\cite{KKS} is 
shown to be zero (the two-loop integrand vanishes point-wise) in Ref.
\cite{KKS}, 
and similarly for the models given in this paper. We shall 
demonstrate the point-wise vanishing of the 
two-loop integrand, using the above defined modular transformation. 

{}We also give an argument for the vanishing
of the higher loop vacuum amplitudes. The argument relies on the
assumption that even for higher loops, we can always go to the unitary gauge
(at least locally on the Riemann surface)
in which the combined contribution of the ghost and longitudinal 
components is spin structure independent. 
This may be achieved with the special choice of $z_i-z_{i+1}=0$.
Again, the higher-loop integrand 
vanishes point-wise in the moduli space. This is the basis of our 
conjecture that the cosmological constant vanishes to all orders in 
the perturbation expansion.

\subsection*{Vanishing of the Vacuum Amplitude}

{}In the following, we give an argument for the vanishing of the two-loop
amplitude.
For our considerations,
we only need to worry about diagrams that involve both $f$ and $g$
twists. Other diagrams which involve only one type of ${\bf Z}_2$ twist
do not break supersymmetry and hence give no contribution to the
cosmological constant. One can show that by multi-loop
modular transformation, any twist that breaks space-time
supersymmetry can be brought to one of the canonical forms:
$(f,g,1,1,\dots,1,1)$ and $(f,1,g,1,1,1,\dots,1,1)$ where the twists
are around the
$(a_1,b_1,\dots,a_g,b_g)$ cycles.
We will show in the following that 
for twists of the first type , the two-loop contribution vanishes
(before summing over the spin structures) because of
the fermionic zero modes -- at least one of the $20$ worldsheet fermions
has odd spin structure, whose theta function is identically zero. 
For twists of the second type, part of the
two-loop contribution vanishes because of the fermionic zero modes, and
the remaining contributions vanish 
(after summing over the spin structures of the first torus)
due to the $(2,0)$ supersymmetry (the supersymmetry that remains after
only the $f$ twist).

{}Let us first consider the twist of the first type
{\em i.e.}, the $f$ and $g$ twists are around the $a$- and $b$-cycle 
respectively of
only one of the tori. 
The worldsheet fermions can naturally be divided into four groups:
\begin{eqnarray}
&(i)~ \Psi_L^{i} \quad \quad   & \mbox{for} ~i=0,1,4,7 \nonumber\\
&(ii)~ \Psi_L^{j} \quad \quad  & \mbox{for} ~j=2,3,5,6,8,9 \nonumber \\ 
&(iii)~ \Psi_R^{i} \quad \quad  & \mbox{for} ~i=0,1,4,7 \nonumber \\
&(iv)~ \Psi_R^{j} \quad \quad & \mbox{for} ~j=2,3,5,6,8,9.
\end{eqnarray}

{}In each of the above groups, the worldsheet fermions have the same 
boundary conditions 
on the tori where the $f$ and $g$ twists do not act
(since the only generating vectors are $V_0$, $V_1$ and $V_2$ on these
tori).
As a result, the fermions in the same group have the same
$(g-1)$-loop spin structure,
{\em i.e.}, they all have odd spin structure or even spin structure.

{}One the other hand, one can easily show that 
for at least one of the groups, the fermions in the same group must
have opposite spin structure in the torus that
both $f$ and $g$ act. The implication is that no matter what the 
$(g-1)$-loop spin structure of this group of fermions is, we can 
always find fermions within the group such that the total spin structure
is odd. Since the $g$-loop $\theta$ function for odd spin structure is
zero, we conclude that the $g$-loop amplitudes of this type of diagrams
vanish.

{}It remains to show that the contribution of the other type of twists
is zero. To show this, let us consider $z_1 \not=z_2$ but $z_1 \rightarrow
z_2$. In this limit, the
combined contribution of the ghosts and longitudinal compoents does not
have spin structure dependence. The $g$-loop amplitude with $f$ and $g$
twists around the $a$-cycles of two adjacent tori is
given by
\begin{equation}
\sum_{\{\alpha^i, \beta^i \}} C^{\alpha^1 \alpha^2 \cdots \alpha^g}_{\beta^1
\beta^2 \cdots \beta^g} Z^{\alpha^1 \alpha^2 \cdots \alpha^g}_{\beta^1 \beta^2
\cdots \beta^g}
\end{equation}
where $\alpha^1 = V_3 + \sum_{j=0}^2 \alpha_j^1 V_j$, 
$\alpha^2 = V_4 + \sum_{j=0}^2 \alpha_j^2 V_j$, 
$\alpha^i = \sum_{j=0}^2 \alpha_j^i V_j$ for $i=3,4,\dots,g$ and 
$\beta^i = \sum_{j=0}^2 \beta_j^i V_j$ for $i=1,2,\dots,g$.

{}Consider the two characters with the same $\alpha^i$ and
$\beta^i$ except for $\alpha^1$: \\
$\bullet$ ~ $\alpha^1 = V_3 + \alpha_1 V_1 + \alpha_2 V_2$ ~;  \\
$\bullet$ ~ $\alpha^1 = V_3 + V_0 + \alpha_1 V_1 + \alpha_2 V_2$ ~.\\
They have the same theta function dependence 
(as only the roles of $\Psi_L^{1,4}$ and $\Psi_L^{0,7}$ are
interchanged). The corresponding phases 
$C^{\alpha^1 \alpha^2 \cdots \alpha^g}_{\beta^1
\beta^2 \cdots \beta^g}$ differ by a factor of 
$-e^{2 \pi i ( \beta_0^1 + \beta_1^1 ) k_{00}}$.
If $k_{00}=0$ or $\beta_0^1 + \beta_1^1 =0$ then the two terms in the $g$-loop
amplitude have opposite signs and so they cancel.
If $k_{00}=1/2$ and $\beta_0^1 + \beta_1^1 = 1$,
the two terms do not cancel. In this case, however,
the fermions in group $(i)$, {\em i.e.}, $\Psi_L^{0,1,4,7}$, have
periodic boundary conditions in the $b$-cycle of the first torus.
On the other hand, around the $a$-cycle of the first torus,
some of the fermions in group $(i)$ have periodic boundary conditions
and some have anti-periodic
boundary conditions. As a result, there are fermions with odd and even
spin structure in the first torus. Therefore,
no matter what the spin structure of the $(g-1)$-loop amplitude is,
there are always fermions with $g$-loop odd spin structure, and hence
the $g$-loop amplitude vanishes due to the fermionic zero mode.

\section{Summary and Remarks}\label{summary}

{}Duality of non-supersymmetric strings have been studied in the 
literature \cite{others}. We see that the models with the bose-fermi 
degeneracy feature are under better control.
Using free fermionic construction, we have constructed 
a set of non-supersymmetric string models
with the property that the 
number of bosonic and fermionic degrees of freedom are equal at each mass 
level. 
Since the models are constructed as {\em Abelian} orbifolds 
(in contrast to the non-Abelian orbifold in Ref. \cite{KKS}), 
explicit calculations of the couplings and $N$-point amplitudes would be
easier to carry out.
 
{}The vanishing of the cosmological constant in the
models has a natural explanation in string theory--- it is simply
a remnant of the $10$-dimensional supersymmetry that defines 
Type II string theory. The exact form of this stringy symmetry
remains to be identified. Understanding this symmetry
would help in constructing more realistic non-supersymmetric models
in which the cosmological constant vanishes only at an isolated
point of the moduli space (instead of having free moduli as in
the present case).

{}Another motivation for considering non-supersymmetric strings
with vanishing cosmological constant is the recent interests
in TeV scale superstings \cite{TeV}. In this new scenario, gauge
and gravitational couplings can unify at the string scale which can be
as low as a TeV. This means the hierarchy problem is absent and so
supersymmetry is no longer needed. (This is a relief, since 
dynamical supersymmetry breaking in string theory is poorly understood).
It is therefore important to explore non-supersymmetric
models with certain realistic features, for instance, vanishing
(or very small) cosmological constant.
However, a realistic model would require non-Abelian gauge group
which is absent in perturbative Type II string theory.
It would be interesting to work out the D-brane spectrum (which can give rise
to non-Abelian gauge group) and see if
the vanishing of the cosmological constant
still persists in the presence of D-branes.

\acknowledgements

{}We thank Shamit Kachru, Eva Silverstein and especially Zurab Kakushadze 
for very valuable discussions. We also thank Jeffrey Harvey for a
correspondence.
The research was partially supported by the
National Science Foundation. G.S. would also like to thank
Joyce M. Kuok Foundation for financial support.



\begin{table}[t]
\begin{tabular}{|c|l|l|l|}
 Sector & ${\bf Z}_2$ orbifold
        & ${\bf Z}_2^{\prime}$ orbifold
        & ${\bf Z}_2 \otimes {\bf Z}_2^{\prime}$ orbifold \\
\hline 
& & &  \\
        & $G_{ij}$, $B_{ij}$, $\phi$ & $G_{ij}$, $B_{ij}$, $\phi$ 
        & $G_{ij}$, $B_{ij}$, $\phi$  \\
${\bf 0}$        & $U(1)^6 \otimes U(1)^2$ & $U(1)^2 \otimes U(1)^6$
        & $U(1)^2 \otimes U(1)^2$  \\
        & $12$ scalars & $12$ scalars & $4$ scalars \\
\hline
& & & \\
 $V_1 + V_2$  & $U(1)^8$ & $U(1)^8$ & $U(1)^4$ \\
     & $16$ scalars & $16$ scalars & $8$ scalars \\
\hline
& & & \\
 $V_1$         & $2$ gravitinos & --- & --- \\
      & $14$ spinors & $16$ spinors 
          & $8$ spinors \\
\hline
& & & \\
 $V_2$         & --- & $2$ gravitinos & --- \\
     & $16$ spinors & $14$ spinors 
          & $8$ spinors \\
 & & & \\
\hline
\hline
& & & \\
$V_3$       & $64$ scalars & N/A & $32$ scalars \\
\hline
& & & \\
$V_3+V_1$   & $32$ spinors & N/A
            & $16$ spinors \\
& & & \\
\hline
\hline
& & & \\
$V_4$       & N/A & $64$ scalars & $32$ scalars \\
\hline
& & & \\
$V_4+V_2$   & N/A & $32$ spinors
            & $16$ spinors \\
& & & \\
\hline
\hline
& & & \\
$V_3 + V_4$ & N/A & N/A & $16$ scalars \\
\hline
& & & \\
$V_3+V_4+V_1+V_2$ & N/A & N/A  & $U(1)^4$ \\
 &  & & $8$ scalars \\
\hline
& & & \\
$V_3+V_4+V_1$ & N/A & N/A & $8$ spinors \\
& & & \\
\hline
& & & \\
$V_3+V_4+V_2$ & N/A & N/A & $8$ spinors \\
& & & \\
\end{tabular}
\caption{The massless spectrum of Model IA. The ${\bf Z}_2$, 
${\bf Z}_2^{\prime}$ orbifolds are supersymmetric and 
are generated by $f$ and $g$ respectively.
The ${\bf Z}_2 \otimes {\bf Z}^{\prime}_2$
orbifold is non-supersymmetric. Here, N/A applies to the sectors
that are absent in the orbifold, and --- indicates that
the states are projected out.}
\label{IA}
\end{table}

\begin{table}[t]
\begin{tabular}{|c|l|l|l|}
 Sector & ${\bf Z}_2$ orbifold
        & ${\bf Z}_2^{\prime}$ orbifold
        & ${\bf Z}_2 \otimes {\bf Z}_2^{\prime}$ orbifold \\
\hline 
& & &  \\
        & $G_{ij}$, $B_{ij}$, $\phi$ & $G_{ij}$, $B_{ij}$, $\phi$ 
        & $G_{ij}$, $B_{ij}$, $\phi$  \\
${\bf 0}$        & $U(1)^6 \otimes U(1)^2$ & $U(1)^2 \otimes U(1)^6$
        & $U(1)^2 \otimes U(1)^2$  \\
        & $12$ scalars & $12$ scalars & $4$ scalars \\
\hline
& & & \\
 $V_1 + V_2$  & $U(1)^8$ & $U(1)^8$ & $U(1)^4$ \\
     & $16$ scalars & $16$ scalars & $8$ scalars \\
\hline
& & & \\
 $V_1$         & $2$ gravitinos & --- & --- \\
      & $14$ spinors & $16$ spinors 
          & $8$ spinors \\
\hline
& & & \\
 $V_2$         & --- & $2$ gravitinos & --- \\
     & $16$ spinors & $14$ spinors 
          & $8$ spinors \\
 & & & \\
\hline
\hline
& & & \\
$V_3$       & $64$ scalars & N/A & $32$ scalars \\
\hline
& & & \\
$V_3+V_1$   & $32$ spinors & N/A
            & $16$ spinors \\
& & & \\
\hline
\hline
& & & \\
$V_4$       & N/A & $64$ scalars & $32$ scalars \\
\hline
& & & \\
$V_4+V_2$   & N/A & $32$ spinors
            & $16$ spinors \\
& & & \\
\hline
\hline
& & & \\
$V_3 + V_4 + V_0$ & N/A & N/A & $U(1)^4$ \\
& & & $8$ scalars \\
\hline
& & & \\
$V_3+V_4+V_0 + V_1 + V_2$ & N/A & N/A  & $16$ scalars \\
\hline
& & & \\
$V_3+V_4+V_0+ V_1$ & N/A & N/A & $8$ spinors \\
& & & \\
\hline
& & & \\
$V_3+V_4+V_0+ V_2$ & N/A & N/A & $8$ spinors \\
& & & \\
\end{tabular}
\caption{The massless spectrum of Model IB. The ${\bf Z}_2$, 
${\bf Z}_2^{\prime}$ orbifolds are supersymmetric and 
are generated by $f$ and $g$ respectively.
The ${\bf Z}_2 \otimes {\bf Z}^{\prime}_2$
orbifold is non-supersymmetric. Here, N/A applies to the sectors 
that are absent in the orbifold, and --- indicates that the
states are projected out.}
\label{IB}
\end{table}

\begin{table}[t]
\begin{tabular}{|c|l|l|l|}
 Sector & ${\bf Z}_2$ orbifold
        & ${\bf Z}_2^{\prime}$ orbifold
        & ${\bf Z}_2 \otimes {\bf Z}_2^{\prime}$ orbifold \\
\hline 
& & &  \\
        & $G_{ij}$, $B_{ij}$, $\phi$ & $G_{ij}$, $B_{ij}$, $\phi$ 
        & $G_{ij}$, $B_{ij}$, $\phi$  \\
${\bf 0}$        & $U(1)^6 \otimes U(1)^2$ & $U(1)^6 \otimes U(1)^6$
        & $U(1)^6 \otimes U(1)^2$  \\
        & $12$ scalars & $36$ scalars & $12$ scalars \\
\hline
& & & \\
 $V_1 + V_2$  & $U(1)^8$ & --- & --- \\
     & $16$ scalars & --- & --- \\
\hline
& & & \\
 $V_1$         & $2$ gravitinos & --- & --- \\
      & $14$ spinors & ---
          & --- \\
\hline
& & & \\
 $V_2$         & --- & $4$ gravitinos & --- \\
     & $16$ spinors & $28$ spinors 
          & $16$ spinors \\
 & & & \\
\hline
\hline
& & & \\
$V_3$       & $64$ scalars & N/A & $32$ scalars \\
\hline
& & & \\
$V_3+V_1$   & $32$ spinors & N/A
            & $16$ spinors \\
& & & \\
\hline
\hline
& & & \\
$V_4$       & N/A & $U(1)^8$ & $U(1)^4$  \\
            &  & $48$ scalars & $24$ scalars \\
\hline
& & & \\
$V_4+V_2$   & N/A & $32$ spinors
            & $16$ spinors \\
& & & \\
\hline
\hline
& & & \\
$V_3 + V_4$ & N/A & N/A & $32$ scalars \\
\hline
& & & \\
$V_3+V_4+V_1$ & N/A & N/A  & $16$ spinors \\
& & & \\
\end{tabular}
\caption{The massless spectrum of Model IIA. The ${\bf Z}_2$, 
${\bf Z}_2^{\prime}$ orbifolds are supersymmetric and 
are generated by $f$ and $g$ respectively.
The ${\bf Z}_2 \otimes {\bf Z}^{\prime}_2$
orbifold is non-supersymmetric. Here, N/A applies to the sectors
that are absent in the orbifold, and --- indicates that the states are
projected out.}
\label{IIA}
\end{table}

\begin{table}[t]
\begin{tabular}{|c|l|l|l|}
 Sector & ${\bf Z}_2$ orbifold
        & ${\bf Z}_2^{\prime}$ orbifold
        & ${\bf Z}_2 \otimes {\bf Z}_2^{\prime}$ orbifold \\
\hline 
& & &  \\
        & $G_{ij}$, $B_{ij}$, $\phi$ & $G_{ij}$, $B_{ij}$, $\phi$ 
        & $G_{ij}$, $B_{ij}$, $\phi$  \\
${\bf 0}$        & $U(1)^6 \otimes U(1)^2$ & $U(1)^6 \otimes U(1)^6$
        & $U(1)^6 \otimes U(1)^2$  \\
        & $12$ scalars & $36$ scalars & $12$ scalars \\
\hline
& & & \\
 $V_1 + V_2$  & $U(1)^8$ & --- & --- \\
     & $16$ scalars & --- & --- \\
\hline
& & & \\
 $V_1$         & $2$ gravitinos & --- & --- \\
      & $14$ spinors & ---
          & --- \\
\hline
& & & \\
 $V_2$         & --- & $4$ gravitinos & --- \\
     & $16$ spinors & $28$ spinors 
          & $16$ spinors \\
 & & & \\
\hline
\hline
& & & \\
$V_3$       & $64$ scalars & N/A & $32$ scalars \\
\hline
& & & \\
$V_3+V_1$   & $32$ spinors & N/A
            & $16$ spinors \\
& & & \\
\hline
\hline
& & & \\
$V_4$       & N/A & --- & --- \\
\hline
& & & \\
$V_4+V_1$   & N/A & ---
            & --- \\
& & & \\
\hline
\hline
& & & \\
$V_3 + V_4$ & N/A & N/A & $32$ scalars \\
\hline
& & & \\
$V_3+V_4+V_1$ & N/A & N/A  & $16$ spinors \\
& & & \\
\end{tabular}
\caption{The massless spectrum of Model IIB. The ${\bf Z}_2$, 
${\bf Z}_2^{\prime}$ orbifolds are supersymmetric and 
are generated by $f$ and $g$ respectively.
The ${\bf Z}_2 \otimes {\bf Z}^{\prime}_2$
orbifold is non-supersymmetric. Here, N/A applies to the sectors
that are absent in the orbifold, --- indicates 
that the states are projected out.}
\label{IIB}
\end{table}

\begin{table}[t]
\begin{tabular}{|c|l|l|l|}
 Sector & ${\bf Z}_2$ orbifold
        & ${\bf Z}_2^{\prime}$ orbifold
        & ${\bf Z}_2 \otimes {\bf Z}_2^{\prime}$ orbifold \\
\hline 
& & &  \\
        & $G_{ij}$, $B_{ij}$, $\phi$ & $G_{ij}$, $B_{ij}$, $\phi$ 
        & $G_{ij}$, $B_{ij}$, $\phi$  \\
${\bf 0}$        & $U(1)^6 \otimes U(1)^2$ & $U(1)^6 \otimes U(1)^6$
        & $U(1)^6 \otimes U(1)^2$  \\
        & $12$ scalars & $36$ scalars & $12$ scalars \\
\hline
& & & \\
 $V_1 + V_2$  & $U(1)^8$ & --- & --- \\
     & $16$ scalars & --- & --- \\
\hline
& & & \\
 $V_1$         & $2$ gravitinos & --- & --- \\
      & $14$ spinors & ---
          & --- \\
\hline
& & & \\
 $V_2$         & --- & $4$ gravitinos & --- \\
     & $16$ spinors & $28$ spinors 
          & $16$ spinors \\
 & & & \\
\hline
\hline
& & & \\
$V_3$       & $64$ scalars & N/A & $32$ scalars \\
\hline
& & & \\
$V_3+V_1$   & $32$ spinors & N/A
            & $16$ spinors \\
& & & \\
\hline
\hline
& & & \\
$V_4$       & N/A & --- & --- \\
\hline
& & & \\
$V_4+V_1$   & N/A & ---
            & --- \\
& & & \\
\hline
\hline
& & & \\
$V_3 + V_4 + V_0 + V_1 + V_2$ & N/A & N/A & $32$ scalars \\
\hline
& & & \\
$V_3+V_4+V_0 + V_2$ & N/A & N/A  & $16$ spinors \\
& & & \\
\end{tabular}
\caption{The massless spectrum of Model IIC. The ${\bf Z}_2$, 
${\bf Z}_2^{\prime}$ orbifolds are supersymmetric and 
are generated by $f$ and $g$ respectively.
The ${\bf Z}_2 \otimes {\bf Z}^{\prime}_2$
orbifold is non-supersymmetric. Here, N/A applies to the sectors
that are absent in the orbifold, --- indicates 
that the states are projected out.}
\label{IIC}
\end{table}

\end{document}